\documentclass{article}
\usepackage{authblk}
\usepackage{graphicx}  

\usepackage{subfigure}

\usepackage{url}  
\usepackage{stfloats} 
\usepackage{amsmath}   

\usepackage{amssymb}
\usepackage{bm}

\usepackage{overpic}
\usepackage[draft]{changes}
	
\newcommand{\bfs}{{\bf s}}
\newcommand{\bfx}{{\bf x}}
\newcommand{\bfr}{{\bf r}}

\newcommand{\srf}{X(\bfs,t)}
\newcommand{\ssrf}{x(\bfs,t)}
\newcommand{\srfp}{X({\bfs}',t')}

\newcommand{\st}{\bfs,t}
\newcommand{\stp}{{\bfs',t'}}
\newcommand{\stpp}{{\bfs'',t''}}
\newcommand{\Eeq}[1]{\left \langle{#1} \right \rangle_{0}}
\newcommand{\EE}[1]{\left \langle{\, #1} \, \right \rangle}
\newcommand{\bfk}{{\bf k}}
\newcommand{\om}{\omega}
\newcommand{\ko}{{\bf k},\omega}

\newcommand{\cor}{C_{\rmx}(\st;\stp)}
\newcommand{\cors}{C_{\rmx}({\bf r},\tau)}

\newcommand{\rmx}{{\rm x}}
\newcommand{\LL}{{\mathcal L}}
\newcommand{\HH}{{\mathcal H}}
\newcommand{\E}{\mathrm{e}}
\newcommand{\de}{\delta}

\newcommand{\ka}{\kappa}
\newcommand{\bo}{\mathbf{0}}
\newcommand{\Rd}{\mathbb{R}^{d}}
\newcommand{\sign}{\mathrm{sign}}

\newcommand{\Ha}{\mathcal{H}}

\newcommand{\noi}{\noindent}
\newcommand{\la}{\lambda}

\newcommand{\beq}{\begin{equation}}
\newcommand{\eeq}{\end{equation}}

\newcommand{\ra}{\rightarrow}

\newcommand{\km}{{\rm k_{c}}}

\newcommand{\e}{{\eta_1}}

\newcommand{\jm}{{\jmath}}

\DeclareMathOperator\erfc{Erfc}

\DeclareMathOperator\sinc{sinc}
\begin{document}

%
%

\title{Space-Time Models based on Random Fields with Local Interactions\thanks{This manuscript is a preprint of an article submitted for consideration in International Journal of Modern Physics B~\copyright 2015, copyright World Scientific Publishing Company, \texttt{http:/www.worldscientific.com/worldscinet/ijmpb}}} 

\author{Dionissios~T.~Hristopulos,
 \thanks{\texttt{dionisi@mred.tuc.gr.}; Corresponding author}}
\author{Ivi~C.Tsantili,\thanks{ \texttt{itsantili@isc.tuc.gr}}} 

\affil{Geostatistics Laboratory, School of Mineral Resources Engineering, \\Technical University of Crete, Chania, 73100 Greece}

\maketitle

\begin{abstract}
The analysis of space-time data from complex, real-life phenomena  requires the use of flexible 
and physically motivated covariance functions. In most cases, it is not possible to 
explicitly solve the  equations of motion for the fields or the respective  covariance 
functions. In the statistical literature, covariance functions are often  based on 
mathematical constructions.  We propose deriving space-time covariance functions by 
solving ``effective equations of motion'', which can be used as statistical 
representations of systems with diffusive behavior.  
In particular, we propose using the linear response theory to formulate 
space-time covariance functions based on an equilibrium effective Hamiltonian. 
The effective space-time dynamics are then 
generated  by a stochastic perturbation around 
the equilibrium point of the classical field Hamiltonian leading to an associated 
Langevin equation. 
We employ a Hamiltonian which extends the classical Gaussian field theory by including 
 a curvature term and leads to a diffusive 
Langevin equation.  
Finally, we derive new forms of space-time covariance functions. 
\end{abstract}


\section{Introduction}
Covariance functions play a significant role in  the 
analysis of space-time data with geostatistical and machine 
learning methods~\cite{Stein05,Rasmussen06,Cressie11}, in inverse modelling~\cite{Tarantola05}, and 
in data assimilation~\cite{Kalnay03}.  Thus, there is active interest in applications of space-time data 
analysis and the development of new covariance models~\cite{Cressie11,Porcu12}. 
{S}{pace-time} covariance functions commonly used are straightforward extensions of 
purely spatial models (e.g., exponential, Gaussian) and  constructions 
based on linear mixtures~\cite{Ma05}. 
More complex models are  
generated using mathematical arguments, e.g. permissibility
conditions~\cite{Ma08} and simplifying assumptions such as separability of
space and time components without clear physical motivation for 
the functional form or the parameters. Several authors have 
proposed more realistic, non-separable covariance models~\cite{deIaco02,Kolovos04,deIaco13}. 
Simple space-time covariance models  are determined by the variance and  constant correlation length 
and correlation time. In purely spatial cases, 
richer parametric families such as the Whittle-Mat\'{e}rn covariance model~\cite{Fuentes05,Guttorp06} or the 
Spartan covariance functions~\cite{dth03,dthsel07,dth14} offer more flexibility. 
The former is derived from a fractional stochastic 
partial differential equation 
driven by white noise~\cite{Whittle54,Lim09}. The latter is obtained 
from a Gaussian field theory with a curvature term which
 also leads to a fourth-order stochastic partial differential equation~\cite{dth06,dth14}.

There is a long history of ideas from physics that find applications in information 
theory and data analysis, including the maximum entropy formalism developed by Jaynes~\cite{Jaynes57}, 
more recently the Bayesian field theory~\cite{Lemm05}, and machine learning applications of the 
variational approximation~\cite{Hoffman13}. 
The connection between statistical physics and space-time statistics on the other hand 
has not been duly appreciated, albeit the former can provide useful models 
for the analysis of space-time data. 
In statistical physics the literature on space-time
fields and correlation functions is extensive~\cite{Forster90}. 
For example, the Gaussian model of 
classical field theory is  equivalent to a 
Gaussian random field with a specific covariance function~\cite{Kardar07,Mussardo10}. 

A formal difference between 
random fields in statistical field theories and those in
spatial statistics is that the former
are defined by means of local interactions, whereas the
latter are defined by means of covariance matrices~\cite{Christakos92,Cressie93}.  
Both approaches are formally equivalent provided that  the covariance function of the 
local interaction model can be explicitly expressed.  
The interaction-based formalism has several advantages for
parameter estimation, interpolation and simulation which derive from the sparseness of the 
inverse covariance matrix  in the 
local interaction representation~\cite{dthsel09,zuk08b}.  Spartan Spatial Random Fields
(SSRFs) are based on the local interaction framework in the static (time-independent) 
case. SSRFs admit explicit relations for the covariance function in one, two, and 
three dimensions~\cite{dth03,dthsel07,eldth08,dthetal08,dth14}. 
A similar framework was independently proposed by
Farmer~\cite{Farmer07}. More recently, covariance functions similar to SSRF 
were derived  for meteorological applications from polynomials of the 
diffusion operator~\cite{Yaremchuk11,Yaremchuk12}.

In statistical physics, a coarse-grained 
Hamiltonian  determines the relative probabilities of different field configurations 
for near-equilibrium systems. 
Excursions from equilibrium leading to dynamic fluctuations follow  
by perturbing the system with noise. The dynamic response of the system 
is determined by   
a \textit{stochastic partial differential equation},
also known as \textit{Langevin equation}, which can be derived in the framework of \textit{linear
response theory}~\cite{Goldenfeld92}. The correlation functions
 are shown to obey the
\textit{fluctuation-dissipation  theorem}, which connects them
to respective susceptibility functions~\cite{Marconi08}. 
In physics, this approach has been applied to study dynamic critical phenomena~\cite{Hohenberg77,Swift77}. 

The physical insight provided by the the fluctuation-dissipation theorem 
is that the fluctuations of the
unperturbed system contain information about the response 
 to external perturbations that drive the system away from equilibrium. 
 This  formalism, to
our knowledge, has not been applied to generate space-time
covariance functions. Our goal in this paper is to show that 
effective Hamiltonians with local interactions, which are successfully used in spatial data analysis,
 can be extended to space-time random
fields.  This contribution is part of an ongoing effort to transfer  
ideas from statistical physics to space-time data analysis~\cite{dth03,dth06,dthetal08,dthsel07,dthsel09,dth14}.
 Using the theory of linear response, we show that the local interaction 
framework leads to explicit forms 
for new space-time covariance functions. 

The remainder of the paper is structured as follows: 
In Section~\ref{sec:math} we present mathematical background on Gaussian random fields. 
Section~\ref{sec:lirf}  reviews local interaction random fields and demonstrates that the
so-called Spartan spatial random field model 
is derived from the principle of maximum entropy. Section~\ref{sec:dynamic-cor} 
briefly presents  linear response theory 
focusing on the calculation of  
correlation functions. 
In Section~\ref{sec:st-ssrf} we use linear response theory
 to derive  equations of motion for space-time SSRF covariances,  
 and we obtain an explicit equation for the SSRF spectral density.  
 Section~\ref{sec:CovFunLE} gives a standard derivation of space-time covariances  
 from Langevin equations.  This leads to a general equation for the space-time covariance function 
 of fields driven by colored noise and recaptures the results of the previous Section in the 
 SSRF case with Gaussian white noise.  In Section~\ref{sec:explicit-cov} we present 
 explicit equations for SSRF space-time covariance functions. 
 Finally, we present our conclusions in Section~\ref{sec:concl}.

\section{Mathematical Preliminaries}
\label{sec:math}

In the following, we focus on Gaussian space-time random
fields. These can be used to model more complicated distributions
in the Bayesian framework of Gaussian random processes~\cite{Rasmussen06}. 
A \textit{space-time random field (STRF)} $ X(\bfs,t;\omega) \in \mathbb{R}$ 
where  $(\bfs,t) \in  \Rd \times \mathbb{R} $ and  $\omega \in \Omega $ is defined as a
mapping from the probability space $(\Omega,A,P)$ into the space of
real numbers so that for each fixed coordinate $(\bfs,t)$, $ X(\bfs,t;\omega)$
is a measurable function of $\omega$ \cite{Christakos92}. An STRF involves by definition many
possible states indexed by $\omega$ \cite{Christakos92,Yaglom87}. In the following, we drop the 
state index $\omega$ for simplicity; instead, we use the symbol $\omega$ for the cyclical frequency 
in Fourier transforms. 
The states (realizations) of the STRF are denoted by the real-valued scalar functions $\ssrf$. 
In the following, $\Phi\left[\srf \right]$ denotes a functional of the STRF 
$\srf$ that takes unique values for each realization $\ssrf$. 

The expectation over the 
ensemble of STRF states is denoted by the angle brackets $\EE{\cdot}.$ Hence, 
the \textit{covariance function} is given by 
\begin{eqnarray}
\label{eq:corel} \cor & =   &    \EE{ X(\bfs',t') \, X(\st) } - \EE{ X(\bfs',t')} \, \EE{X(\st) }.
\end{eqnarray}
An STRF is called \textit{statistically stationary in the weak sense} 
(or simply stationary for brevity), if its expectation is constant and its 
covariance function depends purely on the spatial and temporal lags, $\bfr$ and $\tau$, respectively. 
For simplicity, since we aim to calculate space-time covariance
functions, we assume a zero-mean STRF, i.e. $\EE{\srf}=0.$ 
The STRF variance will be
denoted by $\sigma^{2}$. Furthermore, a stationary random field is called \textit{statistically isotropic}
if its covariance function depends only on the Euclidean distance but not the direction vector. 
In the following we focus on statistically isotropic STRFs. 

In the spectral domain, we use the wavevector $\bfk$ to denote the \textit{spatial frequency} and $\om$ to denote the 
\textit{cyclic frequency} with respect to time. For a given function $x(\bfs,t)$ with space and time dependence, 
we will use $\tilde{x}(\bfk,t)$ to denote the spatial (i.e., with respect to the space variable) 
Fourier transform and $\tilde{x}(\bfk,\om)$ the full Fourier transform with respect to both space and time. 
The pairs of the direct and inverse Fourier transforms, respectively, are defined as follows: 

\begin{subequations}
\label{eq:ft-def}
\beq
\label{eq:ft-s}
\tilde{x}(\bfk,t) = \mathcal{F}_{\bfs}[x](\bfk,t) = \int_{\Rd} d\bfs \, \E^{-\jmath \, \bfk \cdot \bfr} \, x(\bfs,t),
\eeq
\beq
\label{eq:ift-s}
x(\bfs,t)  = \mathcal{F}_{\bfk}^{-1}[\tilde{x}](\bfs,t) = \frac{1}{(2\pi)^d}\int_{\Rd} d\bfk \, \E^{\jmath \, \bfk \cdot \bfr} \, \tilde{x}(\bfk,t),
\eeq
\beq
\label{eq:ft}
\tilde{x}(\bfk,\om) = \mathcal{F}_{\bfs,t}[x](\bfk,\om) =  \int_{\Rd} d\bfs \, \int_{-\infty}^{\infty} dt \,
\E^{-\jmath \, (\bfk \cdot \bfr + \om \, t)} \, x(\bfs,t),
\eeq
\beq
\label{eq:ift}
x(\bfs,t)  = \mathcal{F}_{\bfk,\om}^{-1}[\tilde{x}](\bfs,t) = \frac{1}{(2\pi)^{d+1}}\int_{\Rd} d\bfk \, \int_{-\infty}^{\infty} d\om \, 
\E^{\jmath \, (\bfk \cdot \bfr + \om \, t ) } \, \tilde{x}(\bfk,\om).
\eeq
\end{subequations}

%
%

%

\section{Local Interaction  Random Fields}
\label{sec:lirf}
Time-independent random fields  represent the long-time, i.e., equilibrium, configurations 
of dynamic variables.  They also model ``quenched randomness'' which characterizes the 
structure of geological media and is a key factor for   
subsurface physical processes~\cite{Christakos92}.  
Static random fields can be defined in terms of a 
``pseudo-energy'' functional $\Ha[x(\bfs)]$ which assigns different ``energy'' levels, and 
subsequently different probabilities, to different 
configurations $x(\bfs)$.   In the following, we will assume that the energy functional  
takes real,  non-negative values. 
The most probable configuration of the spatial random field  minimizes $\Ha[x(\bfs)]$. 

The joint \textit{probability density function} (pdf) of the equilibrium spatial random field $X(\bfs;\om)$ 
is determined  from the functional $\Ha[x(\bfs)]$. 
The joint pdf for the realization (state) $x(\bfs)$ 
 is proportional to $\exp[-\Ha[x(\bfs)]$, i.e., 
\[
 f_{\mathrm{eq}}\left[  x(\bfs)\right] = \frac{1}{Z} \, \E^{- \Ha\left[ x(\bfs) \right]},
 \]  

where the partition function $Z$ is given by the following functional integral~\cite{Gelfand48,Feynman65} 
\[
Z=\int \mathcal{D}x(\bfs) \, \E^{- \Ha\left[  x(\bfs)\right]}.
\] 
The functional integral is the continuum limit of the discrete representation 
$\mathcal{D}x(\bfs) = \lim_{n \ra \infty}\prod_{i=1}^{n} dx_{i}$, where 
the vector $(x_{1}, \ldots, x_{n})^{T}$ represents a discretization of the continuum state $x(\bfs)$ at $n$ points.   
 Note that whereas $Z$ may diverge for $n \ra \infty$, the SRF statistical moments 
 are nonetheless well defined. For example, the covariance is given by the following functional 
integral
\beq
\label{eq:stat-mom}
\EE{X({\bfs}_1) \, X({\bfs}_2)} = \frac{1}{Z} \, \int \mathcal{D}x(\bfs) \, x({\bfs}_1) \, x({\bfs}_2)\,
 \E^{- \Ha[x(\bfs)]}. 
\eeq

In  statistical mechanics, the $\Ha[x(\bfs)]$ 
for a given system is obtained from  kinetic and potential energy terms  that reflect 
the motion and interactions of microscopic constituents. 
In the case of macroscopic random fields $\srf$,  
the dynamics may not be fully  known or solvable. Then,  
 $\Ha[x(\bfs)]$ represents an effective functional that incorporates fictitious interactions 
 defined by means of derivatives of the field realizations. 
 Hence, $\Ha[x(\bfs)]$ may represent a sufficiently flexible heuristic functional 
 and not  a first-principles Hamiltonian.

\subsection{Maximum Entropy Formulation of Spartan Spatial Random Fields}
Based on the concept of the effective energy functional, the 
Spartan spatial random field model was proposed~\cite{dth03,dthsel07,dthsel09,dth14}. This 
model is an extension of the classical Gaussian field theory~\cite{Kardar07} which incorporates 
a square curvature term. The respective energy functional is given by
\begin{eqnarray}
\label{eq:fgccont} 
\Ha_{\mathrm fgc}\left[ x(\bfs)\right] & = & 
\frac{1}{{2\eta_0 \xi ^d }} \int_{\Rd} d\bfs \, \left\{
\left[x(\bfs)\right]^2  + \eta_1 \,\xi^2
\left[ {\nabla x(\bfs)} \right]^2  
 + \xi^4 \, \left[ {\nabla^2 \, x(\bfs)} \right]^2 \right\}.
\end{eqnarray}
In the above, $\eta_0$ represents a scale coefficient, $\eta_1$ a rigidity coefficient, and $\xi$ 
a characteristic length; $\nabla$ stands for the gradient and $\nabla^2$ for the Laplace differential operator. 
The energy functional $\Ha_{\rm fgc}\left[ x(\bfs)\right]$ generates 
Gaussian, zero-mean,  stationary random fields. As in classical field theory, a
frequency cutoff is required for differentiability of the field states~\cite{Zinn04}. 
In the absence of such a cutoff, the high-frequency contributions lead to random fields that 
are mean-square continuous but non-differentiable in $d=2,3$~\cite{dthsel07}. 
Explicit expressions for the covariance functions 
in $d=1,2,3$ at the limit of infinite cutoff 
are derived in~\cite{dthsel07,dth14}. The use of~\eqref{eq:fgccont} 
in spatial data analysis requires the estimation of the coefficients $\eta_0$, $\eta_1$ and $\xi$ from 
the data.

The pdf with energy functional~\eqref{eq:fgccont} can be derived 
using the principle of \textit{maximum entropy}~\cite{Jaynes57}. 
The main idea underlying maximum entropy 
is that the pdf is determined by maximizing the entropy of the system given 
 constraints imposed by the data. Furthermore, it is assumed that the data constraints 
are equal to the respective expectations over the pdf.  

Let us  define  the integrals $S_{i}[x(\bfs)$, $i=0,1,2$,  by
\begin{subequations}
\label{eq:constraints}
\begin{align}
{ S_{0}\left[ x(\bfs)\right]} := &  {\int d \bfs \, x^{2}(\bfs)},
\\
{ S_{1}\left[ x(\bfs)\right]} := &   {\int d \bfs \, [ \nabla x(\bfs) ]^{2}},
\\
{ S_{2}\left[ x(\bfs)\right]} := &  {\int d \bfs \, [\nabla^{2} x(\bfs)]^{2}},
\end{align}
\end{subequations}
In addition, we assume that the 
expectations $\EE{ S_{i}[x(\bfs)]}$ can be estimated from the data. 
If we possess only one realization, e.g., $x^{\ast}(\bfs)$, the above assumption requires that the system is ergodic with respect to the constraints, 
i.e., $\EE{ S_{i}[x(\bfs)]} = S_{i}[x^{\ast}(\bfs)]$. 
Then, the maximum entropy pdf conditional on these constraints is given by
\beq
\label{eq:f-max-ent}
f_{\mathrm{eq}}\left[ x(\bfs)\right] = \E^{-\mu - \la_{0} S_{0}\left[ x(\bfs)\right] - \la_{1} S_{1}\left[ x(\bfs)\right] - \la_{2} S_{2}\left[ x(\bfs)\right]}.
\eeq
The constant $\mu$ normalizes the pdf, i.e., $Z = \exp(\mu)$, whereas the constants $\la_{i}$, $i=0,1,2$ 
are in principle obtained by solving the system of the three equations $\EE{ S_{i}[x(\bfs)]} = S_{i}[x(\bfs)]$. 
Note that~\eqref{eq:f-max-ent} is equivalent to~\eqref{eq:fgccont} if 
$\la_0 = {1}/{2\eta_{0} \xi^d}$, $\la_1 = {\eta_{1} \xi^{2-d}}/{2\eta_{0}}$, and $\la_{2} = {\xi^{4-d}}/{2\eta_{0}}$. 


For data sampled on discrete supports, the $S_{i}\left[ x(\bfs)\right]$ are replaced by discretized estimators 
$\hat{S}_{i}(\bfx^{\ast})$, that involve  the sample values 
$(\bfx^{\ast})^{T} =(x_{1}^{\ast}, \ldots x_{N}^{\ast})$.  
For example, on a regular grid the derivatives are 
replaced by respective finite differences which thus implement the high-frequency cutoff~\cite{dth03}.
 On irregular grids, 
estimators can be constructed using kernel functions~\cite{dthetal08,zuk09c}.  
Then, the estimation of the model parameters is reduced to the solution of the following moment equations
\begin{subequations}
\label{eq:solve-constraints}
\begin{align}
\hat{S}_{0}(\bfx^{\ast}) = &   \int_{-\infty}^{\infty}  dx_{1} \ldots \int_{-\infty}^{\infty}  dx_{N} \, \hat{S}_{0}(\bfx) \, \E^{-\mu - \la_{0} \hat{S}_{0}(\bfx) - \la_{1} \hat{S}_{1}(\bfx) - \la_{2} \hat{S}_{2}(\bfx)},
\\
\hat{S}_{1}(\bfx^{\ast}) = &  \int_{-\infty}^{\infty}  dx_{1} \ldots \int_{-\infty}^{\infty}  dx_{N} \, \hat{S}_{1}(\bfx) \, \E^{-\mu - \la_{0} \hat{S}_{0}(\bfx) - \la_{1} \hat{S}_{1}(\bfx) - \la_{2} \hat{S}_{2}(\bfx)},
\\
\hat{S}_{2}(\bfx^{\ast}) = &  \int_{-\infty}^{\infty}  dx_{1} \ldots \int_{-\infty}^{\infty}  dx_{N} \, \hat{S}_{2}(\bfx) \, \E^{-\mu - \la_{0} \hat{S}_{0}(\bfx) - \la_{1} \hat{S}_{1}(\bfx) - \la_{2} \hat{S}_{2}(\bfx)}.
\end{align}
\end{subequations}

\section{Dynamic Correlations}
\label{sec:dynamic-cor}

If a perturbation drives the system away from the equilibrium 
and  the deviation is small, the equilibrium fluctuations 
determine the non-equilibrium response. This idea is exploited in statistical mechanics 
by means of the linear response  theory~\cite{Hohenberg77}.  

 \subsection{Susceptibility Function}

The theory of linear response  focuses on small perturbations 
of the field $\srf$ around the equilibrium state which are caused by an 
external field $h(\bfs,t)$. If we define by $\Eeq{\cdot}$ the 
expectation over the equilibrium distribution and by  $\EE{\cdot}$ the non-equilibrium expectation, the response of
the system to the  external field can be expressed as follows in terms of the susceptibility function 
$\chi(\bfs - \bfs'; t - t')$ and the following convolution equation

\begin{equation}
\label{eq:noneq-lr-1} \EE{\srf} = \Eeq{\srf} + \int_{-\infty}^{t} dt'
\int_{\Rd} d{\bfs}' \, \chi(\bfs - \bfs'; t - t') \, h(\bfs',t').
\end{equation}

\noi For large perturbations, the response should also include nonlinear terms.  

Since the difference $\EE{X(\st)} - \Eeq{X(\st)} = \EE{X'(\st)} $ corresponds to
 the expectation of the 
non-equilibrium fluctuation, Eq.~\eqref{eq:noneq-lr-1} implies that the susceptibility 
is given by the following functional derivative
\begin{equation}
\label{eq:suscept-real} {\chi}(\bfs-\bfs',t -t') = \lim_{h(\stp) \rightarrow 0}
\frac{\delta \EE{{X'}(\st)}}{\delta {h}(\stp)}.
\end{equation}
In the above, $\de (\cdot)/ \de h(\stp)$ represents the functional derivative, which is defined by 
means of $\de h(\st)/\de h(\stp) = \de(\bfs-\bfs')\, \de(t-t')$. 
The limit $h(\stp) \ra 0$ is taken in order to ensure that nonlinear terms vanish. 

Let $\tilde{X'}(\ko)$ and $\tilde{h}(\ko)$ represent the Fourier transforms of the 
STRF fluctuation and the external field which are given by~\eqref{eq:ft-def}. 
The Fourier transform of the susceptibility function is then given by the following limit
\begin{equation}
\label{eq:suscept} \tilde{\chi}(\ko)=\lim_{\tilde{h} \rightarrow 0}
\frac{\delta \EE{\tilde{X'}(\ko)}}{\delta \tilde{h}(\ko)}.
\end{equation}

The susceptibility function is used in physical applications to describe the 
system's response to measurable  external fields (e.g., electric or magnetic fields). 
In the analysis of spatial data, however, 
the external field does not necessarily represent a physical reality; hence, the 
notion of susceptibility has not found wide applicability. Nevertheless, we suggest that there are potential applications of 
the susceptibility concept, 
since the analysis of spatial fields often involves auxiliary variables~\cite{Wackernagel03}, which essentially 
represent external fields. For example,  the elevation is an auxiliary variable that 
has significant impact on rainfall; thus, it makes sense to consider the susceptibility of the rainfall 
field to the elevation of a particular site.

 \subsection{Langevin Equations}
 
Within the framework of linear response theory, the STRF dynamics is  derived from the equilibrium 
energy functional and the noise field that drives deviations from the equilibrium by means of the following 
Langevin equation~\cite{Itzykson89,Goldenfeld92} 

\begin{equation}
\label{eq:noneq} 
\frac{\partial \srf}{\partial t}= - \Gamma \,
\left. \frac{\delta \Ha[x(\bfs)]}{\delta x(\bfs)} \right|_{x(\bfs)= \srf} + \zeta(\st), 
\end{equation}

\noi where $\Gamma$ is a diffusion coefficient, $\delta[.]/\delta x(\bfs)$
denotes the functional derivative with respect to the field state, and $\zeta(\bfs,t)$ is  the 
noise field. The latter is typically Gaussian
white noise with $\EE{\zeta(\st)}=0$ and variance equal to $D$, i.e., 
\begin{equation}
\label{eq:gwn-cov} \EE{\zeta(\st) \, \zeta(\stp)} = D \,
\delta(\bfs-\bfs') \, \delta(t-t').
\end{equation}

Equation~\eqref{eq:noneq} links the rate of change of the 
field state $x(\bfs,t)$ to an  equilibrium-restoring velocity that depends on  $\Ha[x(\bfs)]$ 
 and a random velocity given by the noise term. 
 For example, if $\Ha[x(\bfs)]$  is given by the Ginzburg-Landau effective action, then~\eqref{eq:noneq}
 is known as the time-dependent  Ginzburg-Landau equation~\cite[p.192]{Kardar07}. 
 In the following, we will denote the \textit{restoring velocity} for a given random field state by 
\beq
\label{eq:V}
V[x(\bfs)]:= - \Gamma\, \frac{\delta \Ha[x(\bfs)]}{\delta x(\bfs)},
\eeq
and $V[\srf]$ will denote the respective functional of the random field $\srf$. 

\subsection{Fokker-Planck Equation} 
The pdf of the STRF $\srf$ that is governed by the Langevin equation~\eqref{eq:noneq} 
is the solution of the following \textit{Fokker-Planck} equation~\cite{Goldenfeld92}
\beq
\label{eq:FP}
\frac{\partial f\left[ x(\bfs,t) \right]}{\partial t}= 
\int_{\Rd} d{\bfs}' \, \frac{\delta}{\delta x(\bfs,t) } 
\left[ - V\left[ x(\bfs',t)\right]\, f_{X}\left[ x(\bfs,t) \right] + 
\frac{D}{2} \frac{\delta f_{X}\left[ x(\bfs,t) \right]}{\delta x({\bfs}',t)} \right]
\eeq
The equilibrium (time-independent) pdf is the asymptotic limit (as $t \ra \infty$) 
of the solution of the above \textit{Fokker-Planck} equation:
\begin{equation}
\label{eq:pdf-eq} f_{\mathrm{eq}}\left[x(\bfs)\right] \propto \exp \left\{ - \frac{2 \Gamma \,
\Ha \left[x(\bfs)\right]}{D}\right\}.
\end{equation}

The Fokker-Planck equation may not admit an explicit solution; however, 
for many applications in spatial data analysis it is sufficient to know the covariance of the random field, 
since the fluctuations often follow the Gaussian law, whereas in other cases the data 
can be transformed by means of nonlinear transformations to approximately 
fit the Gaussian law~\cite{Wackernagel03}.

\subsection{Equation of Motion for the Covariance}
To obtain the equation of motion (EOM) for the covariance function,
 we follow the approach described in~\cite[pp.~120-121]{Marconi08}. 
 First, we assume that  $t>t'$ without loss of generality. 
 We use the covariance definition~\eqref{eq:corel}, we replace the time derivative 
 of $\srf$ with~\eqref{eq:noneq}, and then replace $x(\bfs,t)$ with the random fields $\srf$. 
 These steps lead to the following equation
\begin{equation}
\label{eq:cor-fun1} \frac{\partial \cor}{\partial t} = \EE{ \left[
 V[\srf] + \zeta(\st;\om) \right]
\, \srfp}.
\end{equation}

The right hand side of the above equation involves the term  $\EE{\zeta(\st;\om) \,\srfp }$. 
The principle of causality, however, 
 implies that the noise  at time $t$  can not  influence the field at the earlier time $t'$. 
Hence, noise-field cross correlation is dropped, and we obtain the following EOM for the covariance:
\begin{equation}
\label{eq:cor-fun2} \frac{\partial \cor}{\partial t} =  \EE{ V[\srf]  \, \srfp}.
\end{equation}
 

Let us now assume that  $t\leq t'$ then since we can not reason that causality causes the 
noise-field cross correlation to vanish the EOM for the covariance will be given by:

\begin{equation}
\label{eq:cor-fun4} \frac{\partial \cor}{\partial t} =  \EE{ V[\srf]  \, \srfp} +      \EE{  \zeta(\st;\om)  \, X(\stp;\om)}.
\end{equation}

 Moreover  from \eqref{eq:cor-fun2} the partial derivative of the covariance with respect to $t'$ will be given by:

\begin{equation}
\label{eq:cor-fun1} \frac{\partial C(\bfs,t;\bfs',t')}{\partial t'} =  \EE{ V[\srfp]  \, \srf}.
\end{equation}

Next, we  subtract each side of the equation for $\partial \cor /\partial t'$
from the respective side of~\eqref{eq:cor-fun4}  for $\partial \cor /\partial t$. At this point we use the stationarity property,
i.e., that the covariance is a function only of the lag $t -t '$, which implies 
$\partial \cor /\partial t' = -\partial \cor /\partial t.$
These operations lead to the following equation for the covariance rate
\begin{align}
\label{eq:cor-fun-diff} 2\, \frac{\partial \cor}{\partial t}   = &
  ~ \EE{ V[\srf]  \, X(\stp;\om)} \nonumber
\\  &   -             \EE{ V[\srfp] \, X(\st;\om)} \nonumber
\\  &  +      \EE{  \zeta(\st;\om)  \, X(\stp;\om)}.
\end{align}

Using time and space translation invariance, the first two averages on the left hand side
cancel each other. Combining Equations \eqref{eq:cor-fun2}, \eqref{eq:cor-fun4} and \eqref{eq:cor-fun-diff} we obtain 

\begin{equation}
\label{eq:cor-fun1} \frac{\partial C(\bfs,t;\bfs',t')}{\partial t} = \sign(t-t') \EE{ V[\srfp]  \, \srf}.
\end{equation}

\subsection{The Fluctuation-Dissipation theorem}
The \textit{fluctuation-dissipation theorem} links the covariance and the susceptibility 
functions~\cite{Hohenberg77}. For $t>t'$ from \eqref{eq:cor-fun-diff} it is straightforward that: 
\begin{align}
\label{eq:cor-fun-diff_2} 2\, \frac{\partial \cor}{\partial t}   =    \EE{  \zeta(\stp;\om)  \, X(\st;\om)}.
\end{align}

The term in the right hand side of  can be evaluated by means of the 
\textit{Furutsu-Novikov theorem}~\cite{Furutsu63,Novikov65}. The latter states that if $\zeta(\st;\om)$ 
is a Gaussian process and $\phi [\srf]$ is a function of $\srf$ then the following identity holds
\begin{align}
\EE{\phi[\srf] \, \zeta(\stp;\om)}   =   &
 \int d{\bfs}'' \int dt'' \, \EE{ \frac{\delta \phi[\srf]}{\delta \zeta(\stpp;\om)}} 
 \,\EE{\zeta({\bfs}',t';\om) \, \zeta({\bfs}'',t'';\om)}
\end{align}

In the present case, $\phi[\srf]= \srf$. We  use the definition of the
susceptibility function~\eqref{eq:suscept-real} as a response to the noise field, 
and we take into account the noise
covariance~\eqref{eq:gwn-cov}. Then, the time derivative of the 
covariance function is related to the susceptibility function via
\begin{equation}
\label{eq:fluc-diss} 2\, \frac{\partial \cor}{\partial t} =  D \, \chi(\st,\stp).
\end{equation}

Furthermore, for statistically stationary and homogeneous conditions, the above equation becomes
\begin{equation}
\label{eq:fluc-diss-stat} 2\, \frac{\partial \cors}{\partial \tau} = D \, \chi(\bfr,\tau).
\end{equation}

\section{SSRF-based  Space-Time Covariance Functions}
\label{sec:st-ssrf}

Below we derive a partial differential equation for the space-time covariance function based on the equation 
of motion~\eqref{eq:cor-fun2}. 
 We consider a slightly modified form of the Spartan energy functional~\eqref{eq:fgccont}, in which the 
curvature term is multiplied by the coefficient $\mu \ge 0$. In the following, we 
assume an infinite spectral cutoff for the spatial frequency (wavenumber).

We assume that $\Gamma=D/2$ in line with the relaxation \textit{model A} of 
Hohenberg and Halperin~[Eq.(4.2), p.~445] in \cite{Hohenberg77}. 
It can then be shown using the properties of functional derivatives and ignoring boundary effects~\cite{dth06}
that  the Spartan energy
functional is associated with the following restoring velocity 
\begin{equation}
\label{eq:g}
V[\srf] = -\frac{D}{2 \xi^{d} \,
\eta_{0}} \, \left(  1 - \eta_{1} \xi^{2}  \nabla^{2} + \mu \, \xi^{4}
\nabla^{4} \right) \, \srf.
\end{equation}

In light of the above restoring velocity and~\eqref{eq:cor-fun2}, the stationary covariance function 
is given by the solution of  
 the following equation of motion 
\begin{equation}
\label{eq:cor-eom} \frac{\partial \cors}{\partial \tau} = - \tilde{D} \,\sign(\tau) \, 
\left(  1 - \eta_{1} \xi^{2}  \nabla^{2}_{\bfs} + \mu\,\xi^{4}
\nabla^{4}_{\bfs} \right) \, \cors,
\end{equation}
where  $\tilde{D} =\frac{D}{2 \xi^{d}  \, \eta_{0} }$ 
is the combined diffusion coefficient. 
 
 We define by $\tilde{C}(\bfk,\tau)$ the \textit{Fourier transform} of the covariance function
  over the spatial lag vector. 
 This is given by the following $d$-dimensional integral
 \begin{equation}
\tilde{C}(\bfk,\tau) = \mathcal{F}_{\bfr}[C](\bfk,\tau)= \int_{\Rd} d\bfr  \, \E^{-\jmath \bfk \cdot \bfr} \, C(\bfr,\tau),
\end{equation}

The function $\tilde{C}(\bfk,\tau)$ is the spectral density of the covariance at time lag $\tau$. 
Expressing the equation of motion~\eqref{eq:cor-eom} in Fourier space, we obtain the following spectral counterpart  
\begin{equation}
\label{eq:eom-f-fgc-cor}
\frac{\partial \tilde{C}(\bfk,\tau)}{\partial \tau}= - \tilde{D} \, \sign(\tau) \,
\left(  1 + \eta_{1} k^2\xi^{2}  + \mu\, k^4\xi^{4}\right) \, \tilde{C}(\bfk,\tau),
\end{equation}
where $k = \| \bfk \|$. 
The  first-order ordinary differential equation~\eqref{eq:eom-f-fgc-cor} 
admits the following exponentially damped time evolution of the Fourier modes 
\begin{equation}
\label{eq:eom-ft-fgc-cor} \tilde{C}(\bfk,\tau) = 
\tilde{C}(\bfk,0) \, \E^{- \tilde{D}\,\vert\tau\vert \, \left(1 + \eta_{1} k^{2}
\xi^{2} + \mu\, k^{4} \xi^{4} \right) },
\end{equation}
where $\tilde{C}(\bfk,0)$ is the covariance spectral density at zero time 
lag.

Let us  assume statistical isotropy of the spectral density at zero time lag, i.e., $\tilde{C}(\bfk,0)=\tilde{C}(k,0)$. 
Then, based on the \textit{spectral representation}, the space-time covariance function is given in real
space  by the following one-dimensional integral~\cite{Yaglom87}
\begin{align}
\label{eq:fgc-spectral} 
C(\bfr,\tau)  = &
\frac{r^{1-d/2}\,\,\E^{-\tilde{D}\,\vert \tau \vert\,}}{(2\pi)^{d/2}} \, \int_{0}^{\infty} dk \,
\tilde{C}(k,0) \, k^{d/2} \, J_{d/2-1}(kr)  \,\E^{-
\tilde{D}\,\vert \tau \vert\, \left(\eta_{1} k^{2} \xi^{2} + \mu \, k^{4} \xi^{4}
\right) },
\end{align}
where $J_{d/2-1}(x)$ is the Bessel function of the first kind and of order $d/2-1$ . Given that the covariance function is stationary, the expression~\eqref{eq:fgc-spectral} 
 also  holds for $t \ra \infty$. At zero lag, therefore, $C(\bfr,0)$ should be identical to the spatial covariance 
 function that corresponds to the static energy functional~\eqref{eq:fgccont}. 
By setting $\tau =0$  in~\eqref{eq:fgc-spectral}  it follows that the above constraint is satisfied if 
$\tilde{C}(k,0)$ is 
\beq
\label{eq:ssrf-spd}
\tilde{C}(k,0) = \frac{\eta_{0}  \, \xi^d}{1 + \eta_{1}(k \xi)^2 + \mu (k \xi)^4 },
\eeq
which represents the SSRF spectral density given by~\cite{dth03} slightly modified by the presence of  $\mu$. 
According to \textit{Bochner's theorem}~\cite{Bochner59},~\eqref{eq:ssrf-spd} is a permissible 
spectral density if (i) $\tilde{C}(k,0) \ge 0$ and (ii) $\int d\bfk \, \tilde{C}(k,0) < \infty$. 
These conditions are trivially satisfied if $\e>0$, $\mu>0$ and $ d \le 3$. 


\begin{figure}
\centering
\begin{subfigure}[]{}
\includegraphics[width=0.6\textwidth]{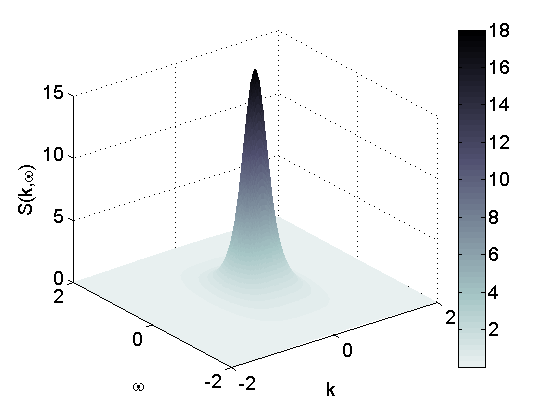}
\label{fig:Spectral density d_1}
\end{subfigure}

\begin{subfigure}[]{}
\includegraphics[width=0.7\textwidth]{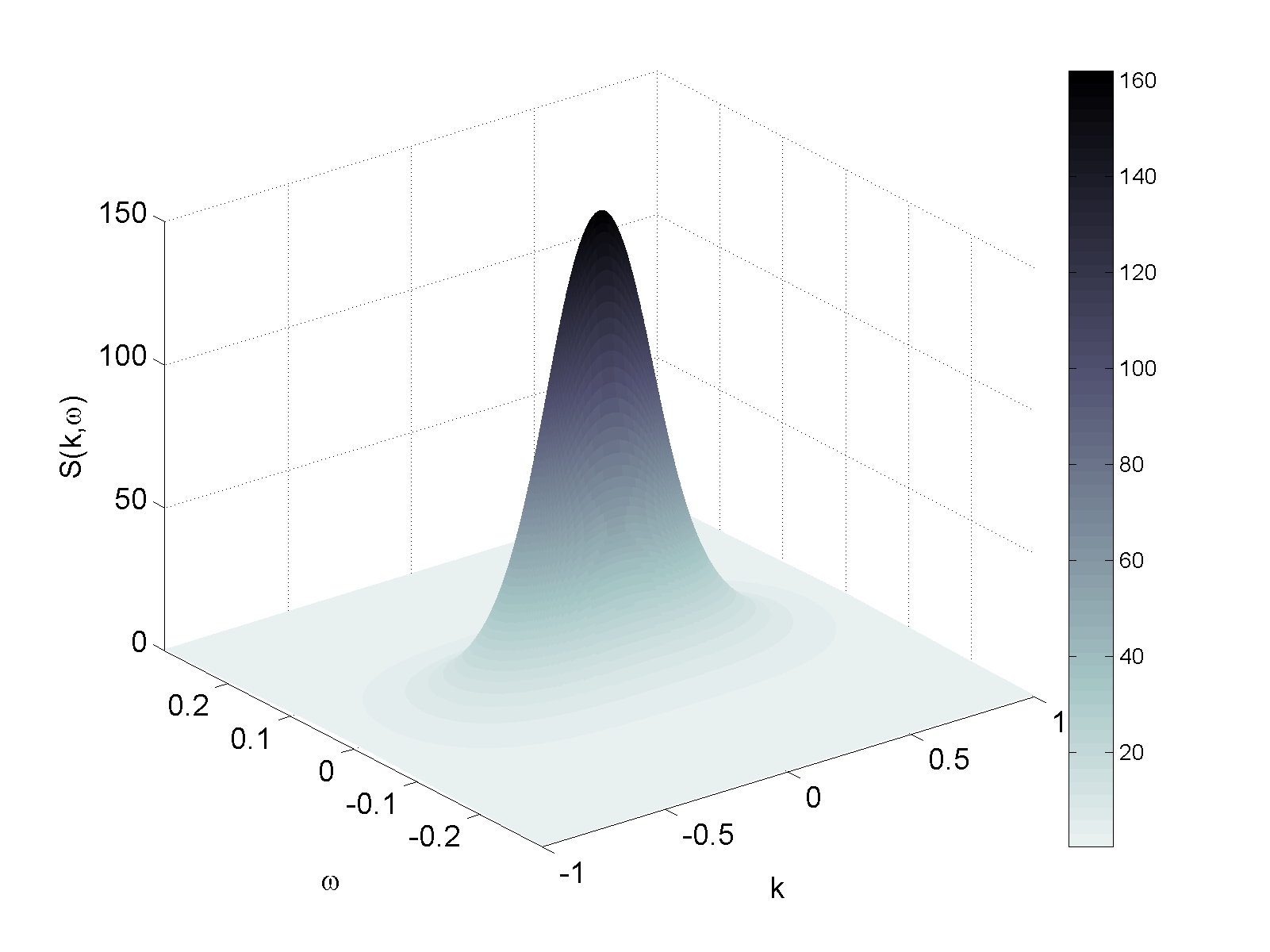}
\label{fig:Spectral density d_1}
\end{subfigure}
\caption{Space-time spectral densities for d=1 (a) and d=3 (b) and for $\e=1, \eta_{0}=1, \xi=3, D=0.5, m=1$ }\label{fig:Spectral density d_1_3}
\end{figure}
The \textit{space-time spectral density} $S(\bfk,\omega) $ is defined as the \textit{Fourier transform} 
of the space-time covariance function in the spatial-temporal frequency domain, i.e:

 \begin{equation}
S(\bfk,\omega)  = \mathcal{F}_{\bfr, \tau}[C](\bfk,\omega)=\int_{-\infty}^{\infty} d\tau \int_{\Rd} d\bfr  \, \E^{-\jmath (\bfk \cdot \bfr+\omega \, \tau)} \, C(\bfr,\tau).
\end{equation}
The spectral density $S(\bfk,\omega)$ can also be expressed by means of the spectral density of the spatial fluctuations at time lag $\tau$
which was obtained in~\eqref{eq:eom-ft-fgc-cor} as follows: 

 \begin{equation}
 \label{eq:Skw}
S(\bfk,\omega)  = \int_{-\infty}^{\infty} d\tau \E^{-\jmath \omega \, \tau} \, \tilde{C}(\bfk,\tau)=
\tilde{C}(\bfk,0) \int_{-\infty}^\infty d\tau \E^{-\jmath \omega \, \tau} \, \E^{- \tilde{D}\,\vert \tau \vert\, \left(1 + \eta_{1} k^{2}
\xi^{2} + \mu\, k^{4} \xi^{4} \right) }.
\end{equation}

Using the isotropic  $\tilde{C}(\bfk,0)$ given by \eqref{eq:ssrf-spd} and taking into account that the 
integrand in~\eqref{eq:Skw} is even, 
$S(\bfk,\omega)$ is expressed in terms of the following  \textit{cosine Fourier transform}

\begin{align}
\label{eq:fgc-spectral density-d-1} 
S(\bfk,\omega) = &
\frac{2\eta_0 \xi}{ 1 + \e k^2\xi^2 + \mu\,k^4 \xi^4 } \, \int_{0}^{\infty} d\tau  \cos(\omega\tau)  \,\E^{-
\tilde{D}\,\vert \tau \vert\, \left(1+\eta_{1} k^{2} \xi^{2} + \mu \, k^{4} \xi^{4}
\right) }.
\end{align}

For $\eta_{1}>0$ and $\tilde{D}>0$  the following analytic expression for the the SSRF space-time spectrum is obtained 
 using the identity [Eq.(1.4.1) p.14] in~\cite[Eq.(1.4.1) p.14]{magnus1954tables}

\begin{align}
\label{eq:fgc-spectral density} 
S(\bfk,\omega) = &
\frac{2 \eta_0 \xi\tilde{D}}{ \tilde{D}^{2}(1 + \e k^2\xi^2 + \mu\,k^4 \xi^4)^{2}+\omega^2} . 
\end{align}

The above is a permissible spectral density function of the form $S(\bfk,\omega)=a/(\Pi(\bfk)+\omega^2)$, where $\Pi(\bfk)$ is an eighth order polynomial. This model is  more flexible than the one described by the  second order Mat\'{e}rn spectral density, commonly used in geostatistics, in which the respective polynomial  $\Pi(\bfk)$  is given by  $\Pi(\bfk)=(\bfk^2+\phi^{2})^{4}$~\cite{jones1997models}. 
In Figure~\ref{fig:Spectral density d_1_3}, the spectral density~\eqref{eq:fgc-spectral density} 
is plotted for $d=1$ and $d=3$  for $\mu=1$. 
The spectral density~\eqref{eq:fgc-spectral density}  could by used in meteorological modelling 
as space-time extension of 
purely spatial spectral densities derived by polynomials of the diffusion operator~\cite{yaremchuk2012multi}.

\section{Covariance Function from Langevin equation}
\label{sec:CovFunLE}

In this section we  obtain space-time covariances
from the  equation that governs the motion of the STRF  realizations. 
Based on~\eqref{eq:noneq} and~\eqref{eq:g} the following Langevin equation governs the 
time evolution of the state $\ssrf$

\beq
\label{eq:langevin-real}
\frac{\partial x(\bfs,t)}{\partial t} = V[x(\bfs,t)] + \zeta(\bfs,t),
\eeq
where $V[x(\bfs,t)]$ is the restoring ``velocity'' as defined by~\eqref{eq:V}, 
 and  $\zeta(\bfs,t)$ is the 
stochastic ``velocity''. 
We assume a bilinear equilibrium energy functional of the form

\[
\HH\left[ x(\bfs)\right]= \int_{\Rd} d{\bfs}\int_{\Rd} d{\bfs}' \, x(\bfs') \, \LL(\bfs' - \bfs) \, x(\bfs) ,
\]
where $\LL(\bfs' - \bfs)$ is the inverse spatial covariance operator defined by means of the 
convolution integral~\cite{Xu05} 
\[ 
\int_{\Rd} d \bfs_{1} \, \LL(\bfs-\bfs_{1}) \, C(\bfs_{1} -\bfs') = \delta(\bfs - \bfs').
 \]
 Note that $\LL(\bfs' - \bfs)$, $\Rd \mapsto \mathbb{R}$ is a \textit{real-valued} and \textit{symmetric} function, 
 i.e., $\LL(\bfs' - \bfs) = \LL(\bfs - \bfs')$. 
The restoring velocity is given by
\[
V[x(\bfs,t)] = - \int_{\Rd}  d \bfs_{1} \,  \LL(\bfs-\bfs_{1}) \, x(\bfs_{1},t).
\]
The Langevin equation corresponding to~\eqref{eq:langevin-real} 
for the spatial Fourier modes of the STRF  is
\[
\frac{\partial \tilde{x}(\bfk,t)}{\partial t} = 
- \widetilde{\LL}({\bfk}) \, \tilde{x}(\bfk,t) + \tilde{\zeta}(\bfk,t),
\]
where $\widetilde{\LL}({\bfk}) \in \mathbb{R}$ is the inverse spatial covariance operator  in Fourier space. 
If $\LL(\cdot)$ is a linear combination of even-order derivative operators, then $\widetilde{\LL}({\bfk})$ 
respectively is a polynomial in $k^2$. 

The temporal evolution of the Fourier modes is given by the solution of the above
ordinary differential equation, i.e., by
\begin{align}
\label{eq:path_functions}
\tilde{x}(\bfk,t)= \tilde{x}(\bfk,0) \, \E^{-\widetilde{\LL}(\bfk)\, t} + \int_{0}^{t} dt' \,
\E^{-\widetilde{\LL}(\bfk)\, (t-t')}\, \tilde{\zeta}(\bfk,t').
\end{align}

The STRF spatial spectral density is given by $\tilde{C}(\bfk,t;\bfk',t') = \EE{\tilde{x}(\bfk,t)\,\tilde{x}(\bfk',t')}_{c}$.
It follows from the above that the temporal evolution of $\tilde{C}(\bfk,t;\bfk',t')$  
is determined by the integral equation
\begin{align}
\label{eq:central_cov}
\EE{\tilde{x}(\bfk,t)\,\tilde{x}(\bfk',t')}_{c}  
= &~\E^{-\widetilde{\LL}(\bfk)\, t}\, 
\E^{-\widetilde{\LL}(\bfk')\, t'} \,  \EE{\tilde{x}(\bfk,0)\,\tilde{x}(\bfk',0)\,}_{c} 
\\\nonumber&+\int_{0}^{t} dt_1 \int_{0}^{t'} dt_2
\E^{-\widetilde{\LL}(\bfk)\, (t-t_{1})}\, 
\E^{-\widetilde{\LL}(\bfk')\, (t'-t_{2})} \,  \EE{\tilde{\zeta}(\bfk,t_1)\,\tilde{\zeta}(\bfk',t_2)\,},
\end{align}

\noi where  $\EE{\tilde{x}(\bfk,0)\,\tilde{x}(\bfk',0)}_{c}$ is the initial condition at zero time lag.

If we assume a correlated  stochastic ``velocity'' with spatial spectral density  
\[
\EE{\tilde{\zeta}(\bfk,t_{1})\,\tilde{\zeta}(\bfk',t_{2})\,}= 2D \,(2\pi)^{d} \, \delta(\bfk+\bfk') \,
 \tilde{c}_{\zeta}(\bfk) \, \delta(t_{1}-t_{2}),
\]
%
%

\noi we obtain the following integral equation for the temporal evolution of
$\tilde{C}(\bfk,t;\bfk',t') $ 
\begin{align}
\label{eq:central_cov_2}
 \tilde{C}(\bfk,t;\bfk',t')  = ~&\E^{- [ \widetilde{\LL}(\bfk) \, t + \widetilde{\LL}(\bfk') \, t'] }
 \, \tilde{C}(\bfk,0;\bfk',0) \\&\nonumber+~ (2\pi)^d \,D  \,  \int_{0}^{\min(t,t')} dt_1
\E^{- [ \widetilde{\LL}(\bfk) \, (t-t_{1}) + \widetilde{\LL}(\bfk') \, (t'-t_{1})] }\, 
 \delta(\bfk +\bfk')\,
 \tilde{c}_{\zeta}(\bfk) 
.
\end{align}
 Setting $t=t'$ in \eqref{eq:central_cov_2} we obtain

\begin{align}
\label{eq:central_cov_3}
 \tilde{C}(\bfk,t;\bfk',t)  = ~&\E^{-  [\widetilde{\LL}(\bfk) + \widetilde{\LL}(\bfk') ]\, t  }\,
 \tilde{C}(\bfk,0;\bfk',0) \\&\nonumber+~ (2\pi)^{d}\,D \,\frac{ \delta(\bfk+\bfk')\,
 \tilde{c}_{\zeta}(\bfk) }{\widetilde{\LL}(\bfk)+\widetilde{\LL}(\bfk')} \, 
 \left[ 1- \E^{- (\widetilde{\LL}(\bfk) + \widetilde{\LL}(\bfk') )\, t  } \right].
\end{align}

\noi Due to stationary  it holds that 
$\tilde{C}(\bfk,t;\bfk',t)= \tilde{C}(\bfk,t';\bfk',t') $ for all $t$ and $t'$.  
Hence, we obtain the following relation for the spatial spectral density at zero time lag
\begin{align}
\label{eq:central_cov_zero}
\tilde{C}(\bfk,0;\bfk',0) =\lim_{t \ra \infty} \tilde{C}(\bfk,t;\bfk',t)
=(2\pi)^{d}\,D \,\frac{ \delta(\bfk+\bfk')\,
 \tilde{c}_{\zeta}(\bfk) }{\widetilde{\LL}(\bfk)+\widetilde{\LL}(\bfk')}.
\end{align}

\noi 
Combining~\eqref{eq:central_cov_2} and~\eqref{eq:central_cov_zero}, 
the temporal evolution of the spatial spectral density is given by the following exponential 
relation

\begin{align}
\label{eq:Cov_fin}
 \tilde{C}(\bfk,t;\bfk',t')  =~ \tilde{C}(\bfk,0;\bfk',0)  \, 
 \E^{- [\theta(t-t') \widetilde{\LL}(\bfk)\,(t-t')+\theta(t'-t) \widetilde{\LL}(\bfk')\,(t'-t)] }.
\end{align}
Finally, in light of~\eqref{eq:Cov_fin} and by invoking the reflection symmetry 
$\widetilde{\LL}(\bfk) = \widetilde{\LL}(-\bfk)$,  
 we obtain the following equation for the evolution of the space-time covariance
  
\begin{align}
\label{eq:zc}
C(\bfs,t;\bfs',t') & =  \int_{\Rd} \frac{d\bfk}{{(2\pi)}^d} 
\int_{\Rd}\frac{d\bfk'}{{(2\pi)}^d} \,
\, \E^{\jmath (\bfk\,\bfs + \bfk'\,\bfs')}\,  
\tilde{C}(\bfk,t;\bfk',t')
\\  \nonumber & = 
\frac{D}{2} \, \int_{\Rd} \frac{d\bfk}{{(2\pi)}^d}\,
 \E^{\jmath \bfk\,(\bfs -\bfs') } \, \E^{- \widetilde{\LL}(k)\,\vert t-t' \vert  }
 \, \left( \frac{\tilde{c}_{\zeta}(\bfk) }
 {\widetilde{\LL}(k)} \right)
.
\end{align}

Equation~\eqref{eq:zc} holds  in general for colored noise and for general $\widetilde{\LL}(k)$. 
For the SSRF case,  the inverse spatial covariance operator $\widetilde{\LL}(k)$ is given by~\cite{dth14}
\[
\widetilde{\LL}(k)= \frac{D}{2\,\eta_0 \xi^d} \left[ 1 + \e k^2\xi^2 + \mu\,k^4 \xi^4 \right].
\] 
Assuming that the stochastic ``velocity'' is Gaussian white noise, i.e., $\tilde{c}_{\zeta}(\bfk)=1$, 
 the  SSRF covariance $C(\bfr,\tau)$ is given by
\begin{align}
C(\bfr;\tau)& =  \int_{\Rd}\frac{d\bfk}{{(2\pi)}^d} \, \E^{\jmath \bfk\,\bfr }   
\,\frac{\eta_0 \xi^d }{ 1 + \e k^2\xi^2 + \mu\,k^4 \xi^4 }\,
\E^{-\frac{D}{2\,\eta_0 \xi^d} \left( 1 + \e \,k^2\xi^2 + \mu\,k^4 \xi^4 \right)\vert \tau \vert}. 
\end{align}
Based on the above and the spectral representation of isotropic random fields which 
transforms the multidimensional integral into a one-dimensional integral~\cite{Yaglom87}, 
we  recover~\eqref{eq:fgc-spectral} derived above.

\section{Explicit Space-time Covariance Functions}
\label{sec:explicit-cov}
Equation~\eqref{eq:fgc-spectral} is the general result for the space-time covariance function 
obtained  in the framework of linear response theory from the equilibrium local-interaction 
energy functional~\eqref{eq:fgccont}.  
We use the following integral identity for the spectral density 
\beq
\label{eq:spd-A}
\frac{\eta_{0} \xi^d}{1 + \e \, k^{2} \,\xi^{2}  + \mu\, k^{4} \,\xi^{4}} = \int_{0}^{\infty} {d \ka} 
\, \E^{-\ka\,\left[ 1 + \eta_{1}(k \xi)^2 + \mu (k \xi)^4 \right]/( \eta_{0}\, \xi^d) }.  
\eeq
The above  is based 
on $1/A = \int_{0}^{\infty} {d \ka} \, \E^{-\ka\,A}$, for $A>0$.

The above leads to a space-time covariance function which is given by 
\begin{subequations}
\label{eq:ssrf-st} 
\begin{align}
\label{eq:ssrf-st-1} 
C(\bfr,\tau)  = &
\frac{r^{1-d/2}\,\E^{-\tilde{D}\,\vert \tau \vert\,}}{(2\pi)^{d/2}} \, \int_{0}^{\infty} d\ka \, 
\int_{0}^{\infty} dk \, k^{d/2} \, J_{d/2-1}(kr)  \, \E^{-A(\ka, k)}, 
 \\
 \nonumber \\
A(\ka, k)  = &
\tilde{D}\,\vert \tau \vert\, \left(\eta_{1} k^{2} \xi^{2} + \mu \, k^{4} \xi^{4}\right) 
    + \ka\,\left[ 1 + \eta_{1}(k \xi)^2 + \mu (k \xi)^4 \right]/( \eta_{0}\, \xi^d).
\end{align}
\end{subequations}
We can thus express the space-time covariance function in terms of the following double integral
\begin{align}
\label{eq:ssrf-st-2}
C(\bfr,\tau)  = & \frac{r^{1-d/2}\,\E^{-\tilde{D}\,\vert \tau \vert\,}}{(2\pi)^{d/2}} \, \int_{0}^{\infty} d\ka \, Q(\ka),
\\
\label{eq:Q1}
Q(\ka) = & 
\int_{0}^{\infty} dk \, k^{d/2} \, J_{d/2-1}(kr)  \, \E^{-A(\ka, k)}.
\end{align}

\subsection{Covariance for the zero-curvature model}

We investigate the space-time covariance~\eqref{eq:ssrf-st}  with $\mu=0$.
In this case the function $A(\ka, k) $ is given by
\beq
\label{eq:A}
A(\ka, k) = \frac{\ka}{\eta_{0} \xi^d} + 
\left( \tilde{D}\,\vert \tau \vert\, \eta_{1}  \xi^{2}  + \frac{\ka \e \xi^{2}}{\eta_{0} \xi^d}\right) k^{2}.
\eeq
Then, the spectral integral~\eqref{eq:Q1} can be analytically performed 
 using~[Eq.(11.4.28), p.~486]~\cite{AbraSte}\
\beq
\label{eq:spec-ssrf-int}
\int_{0}^{\infty} d k \, \E^{-a^2\,k^2} \, k^{\mu-1}\, J_{\nu}(bk) = 
\frac{\Gamma\left( \frac{\nu+\mu}{2}\right)}{2 a^{\mu} \,\Gamma(\nu+1)}\, 
\left( \frac{b}{2a}\right)^{\nu} \, {}_{1}F_{1}\left(\frac{\nu+\mu}{2}, \nu+1; - \frac{b^2}{4a^2}\right),
\eeq

In the above, the ${}_{1}F_{1}(a_1,a_2; z)$ ---also known as $M(a_1,a_2; z)$--- is the 
\textit{confluent hypergeometric function}, defined by~[Eq.~(13.1.2), p.~504]~\cite{AbraSte} 
as follows 
\[
M(a_1,a_2; z)=\sum_{n=0}^{\infty} \frac{(a_{1})^{(n)} z^{n}}{(a_{2})^{(n)}\, n!},
\]

where 
$(a_{1})^{(n)}$ is the rising factorial,  
\[
(a_{1})^{(n)}= a_{1} (a_{1}+1) \ldots (a_{1}+n-1), \quad (a)_{1}^{(0)}=1. 
\]

We apply the above definition with $a^2 = 
\tilde{D}\,\vert \tau \vert\, \e \,  \xi^{2} + \ka\,\e \, \xi^2 /( \eta_{0}\, \xi^d)$, 
$\nu = d/2-1$, $\mu =d/2 +1$, and $b=r$ to the integral in~\eqref{eq:spec-ssrf-int}. 
Based on these replacements,~\eqref{eq:Q1} and~\eqref{eq:A}, we evaluate the function $Q(\ka)$ 
as follows

\begin{align}
\label{eq:Qal}
 Q(\ka)  = &
 \frac{\E^{-\ka/(\eta_{0} \xi^d)}}{2 a^{d/2+1}} \,\left( \frac{r}{2a} \right)^{d/2-1}\, 
{}_{1}F_{1}\left(\frac{d}{2},\frac{d}{2};\frac{-r^2}{4a^2}\right).
\end{align}

Further, using the definitions for the hypergeometric function,  it follows for $a_{1}=a_{2}=d/2$ that 
$M(d/2,d/2; z)=\exp(z).$  In light of this identity, and the function $Q(\ka)$ is given by
\begin{align}
\label{eq:Qal-2} 
 Q(\ka)  = &
 \frac{r^{d/2-1}}{2^{d/2}\, a^d} \, \E^{- \frac{r^2}{4\, a^2}- \frac{\ka}{\eta_{0} \xi^d}}.
\end{align}

Hence, the expression~\eqref{eq:Qal-2} for $Q(\ka)$ and~\eqref{eq:ssrf-st-2} 
lead to the following univariate integral for the space-time covariance function
\begin{subequations}
\label{eq:cov-ssrf-st}
\beq
\label{eq:cov-ssrf-st-1}
C(\bfr,\tau)  =  \frac{\E^{-\tilde{D}\,\vert \tau \vert}}{(4\pi)^{d/2}} \, \int_{0}^{\infty} d\ka \, 
\frac{\E^{- \frac{r^2}{4\, \left( \beta_{1} + \beta_{2} \ka\right) }- 
\ka \beta_{0}}}{\left( \beta_{1} + \beta_{2} \ka\right)^{d/2}},
\eeq
where 
\beq
\label{eq:cov-ssrf-st-2}
\beta_{0} = \frac{1}{\eta_{0} \xi^d}, \; \beta_{1} = \tilde{D}\,\vert \tau \vert\, \left(\eta_{1}  \xi^{2}\right), \; \beta_{2} = 
\frac{ \eta_{1}\xi^2}{  \eta_{0}\, \xi^d}.
\eeq
\end{subequations}

Equations~\eqref{eq:fgc-spectral} and~\eqref{eq:cov-ssrf-st}  both involve univariate integrals. 
The latter, however, involves a non-negative, decreasing function of $\ka$, whereas the former involves the integration of the 
oscillating Bessel function $J_{d/2-1}(kr)$ which is more difficult to evaluate numerically, especially for large lags. 

\subsubsection{The case of zero spatial lag}

For $\bfr = {\mathbf 0}$ the covariance function~\eqref{eq:cov-ssrf-st} is given by the integral 
\beq
\label{eq:cov-ssrf-zero-s}
C(\bo,\tau)  =  \frac{\E^{-\tilde{D}\,\vert \tau \vert}}{(4\pi \beta_{2})^{d/2}} \, \int_{0}^{\infty} d\ka \, 
\frac{\E^{- \ka \beta_{0}}}{\left( \ka +  \frac{\beta_{1}}{\beta_{2}} \right)^{d/2}}.
\eeq

The integral can be performed using integral tables, and more precisely [Eq.~(3.362.2), p.~362]~\cite{Gradshteyn94} 
for $d=1$, 
[Eq.~(3.352.4), p.~358]~\cite{Gradshteyn94}  for $d=2$ 
and [Eq.~(3.369), p.~363]~\cite{Gradshteyn94} for $d=3$ which lead to the 
following expressions

\begin{align}
\label{eq:cov-ssrf-zero-s2}
C(\bo,\tau)  = & \left\{ 
                \begin{array}{c}
               \frac{\eta_{0}}{2} \, \sqrt{ \frac{1}{\e} }\, \erfc\left(\sqrt{\tilde{D}\,\vert \tau \vert}\right),     \\
                    -  \frac{\eta_{0}} {4\pi \e } \, \mathrm{Ei}(-\tilde{D}\,\vert \tau \vert), \\  
                      \frac{2 \eta_{0} \E^{-\tilde{D}\,\vert \tau \vert}}{\sqrt{\tilde{D}\vert \tau \vert}\left( 4\pi \e \right)^{3/2}}
                      - \frac{2 \eta_{0} }{{\pi (4 \e)^{3/2}}}\, 
                         \erfc\left(\sqrt{\tilde{D}\,\vert \tau \vert}\right).
                \end{array}
                \right. 
\end{align}
In the above, $\mathrm{Ei}(x)$ is the \textit{exponential integral function} defined by 
$\mathrm{Ei}(x) = \int_{-\infty}^{x} dt\, \E^{t}/t$ and 
$\erfc(x)$ is the \textit{complementary error function} defined by 
$\erfc(x) = \frac{2}{\sqrt{\pi}}\int_{x}^{\infty} dt \,\E^{-t^2}$. 
The expansion of  $\mathrm{Ei}(x)$ around zero is given by
\[
\mathrm{Ei}(x) = \gamma + \ln|x| + \sum_{k=1}^{\infty } \frac{x^{k}}{k!}
\]
where $\gamma$ is the Euler-Mascheroni constant. In light of the above and~\eqref{eq:cov-ssrf-zero-s2}
it follows that the time covariance has a singularity at $\tau =0$ for $d=2,3$.   

\subsubsection{The case of zero time lag}

 For $\tau=0$ it follows that $\beta_{1}=0$ and the covariance function~\eqref{eq:cov-ssrf-st}
  is given by the following integral 
 
 \beq
\label{eq:cov-ssrf-zero-t}
C(\bfr,0)  =  \frac{1}{(4\pi \beta_{2})^{d/2}} \, \int_{0}^{\infty} d\ka \, 
\frac{\E^{- \frac{r^2}{4\,   \beta_{2} \ka }- \ka \beta_{0}}}{\ka^{d/2}} = 
\frac{\beta_{0}^{d/2}}{(4\pi \beta_{2})^{d/2}\,\beta_{0}} \, \int_{0}^{\infty} dy \, 
\frac{\E^{- \frac{\mu^2}{4\,  \,y} - y }}{y^{d/2}},
\eeq

\noi where $ y = \beta_{0} \ka$ and $\mu = r\sqrt{\beta_{0}/\beta_{2}} = r/\xi\sqrt{\e} $. 
This integral can be evaluated using the identity [3.471.12, p.~385]~\cite{Gradshteyn94} which leads to

 \beq
\label{eq:cov-ssrf-zero-t2}
C(\bfr,0)  =  
\frac{2^{d/2} \eta_{0} }{(4\pi \e )^{d/2}} \,  \left( \frac{r}{\xi \sqrt{\e}} \right)^{1-d/2} \, 
K_{d/2-1}\left( \frac{r}{\xi\sqrt{\e}} \right),
\eeq

\noi where $K_{d/2-1}(\cdot)$ is the modified Bessel function of the second kind of order $d/2-1$. 
For $d=1$ the covariance function depends on $K_{-1/2}(\cdot) = K_{1/2}(\cdot)$, for $d=2$ it depends 
on $K_{0}(\cdot)$, whereas for $d=3$ the function $K_{1/2}(\cdot)$ 
is involved. In both $d=1$ and $d=3$ the behavior of the Bessel function near zero is 
\begin{align*}
K_{\pm 1/2}(x) \sim   & \left\{ 
                \begin{array}{c}
                \frac{\Gamma(1/2)}{2} \left( \frac{2}{x}\right)^{1/2},
                \; d=1, 3
                \\
                -\ln(x/2) - \gamma, \; d=2.
                \end{array}
                \right. 
\end{align*}
Hence, in light of~\eqref{eq:cov-ssrf-zero-t2} the spatial covariance has a singularity at $\bfr = \mathbf{0}$ 
in $d=2, 3$, in agreement with  the limit of the time covariance as well. This is not surprising since 
the model described above is identical to the Gaussian field theory, which is known to have a singular behavior at the 
origin if a finite cutoff is not used~\cite[p.~227]{Mussardo10}.


%

\subsubsection{Explicit covariances for general lags}

\begin{figure}
\centering
        \begin{subfigure}[]{}
                \includegraphics[width=0.45\textwidth]{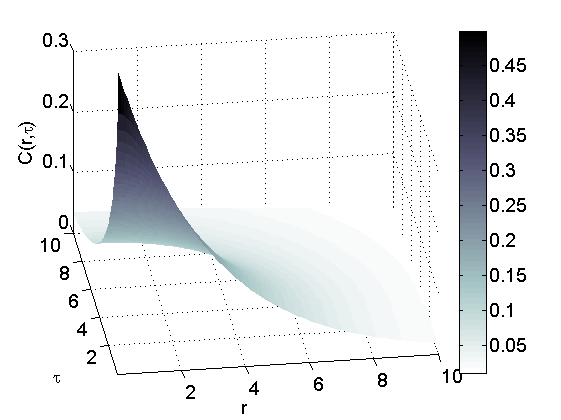}
                \label{fig:ST cov d_1}
        \end{subfigure}
         \begin{subfigure}[]{}
                \includegraphics[width=0.45\textwidth]{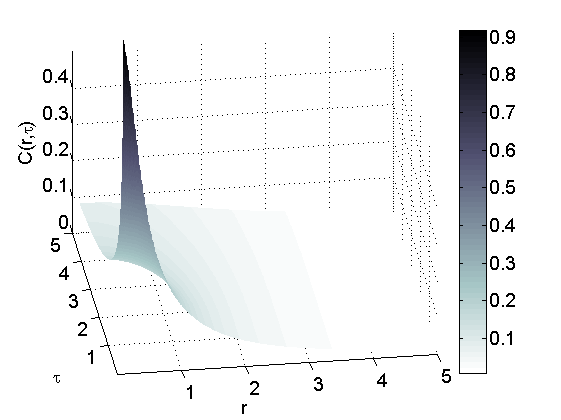}         
               \label{fig:ST cov d_3}
        \end{subfigure}  
 \caption{Space-time covariances for (a) d=1 and (b) d=3. In both cases
  $\e=1, \eta_{0}=1, \xi=3, \tilde{D}=1 $. In (a) the minimum lag ---in both space and time--- is equal 
  to the floating point relative accuracy, whereas 
  in (b) it is equal to $0.1$ to avoid the singularity at $(0,0)$.}\label{fig:ST cov}     
\end{figure}

\begin{figure}
\centering
\begin{subfigure}[]{}
\includegraphics[width=0.45\textwidth]{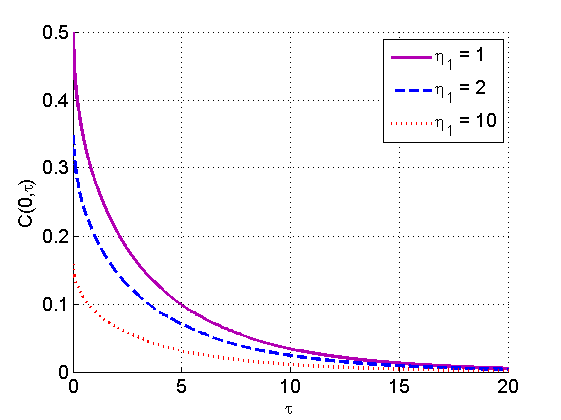}
\label{fig:C2 Timeseries}
\end{subfigure}
\begin{subfigure}[]{}
\includegraphics[width=0.45\textwidth]{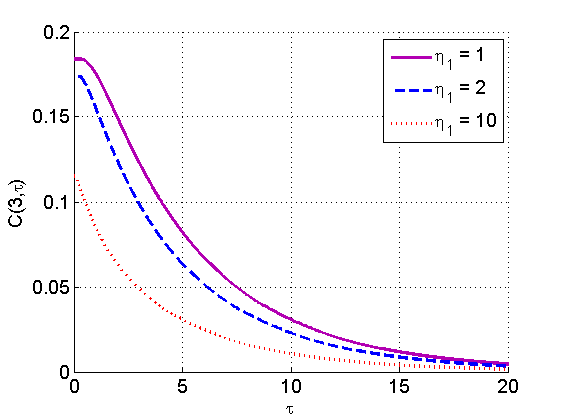}
\label{fig:R7 Timeseries}
\end{subfigure}
\caption{Space-time covariances for $d=1,  \eta_{0}=1, \xi=3, \tilde{D}=1 $. 
(a): Covariances for zero space lag and different $\e$ are shown. 
 (b): Covariances for $r=3$ and different $\e$. }\label{fig:1D_space_time_covariance_t}
\end{figure}

\begin{figure}
\centering
\begin{subfigure}[]{}
\includegraphics[width=0.45\textwidth]{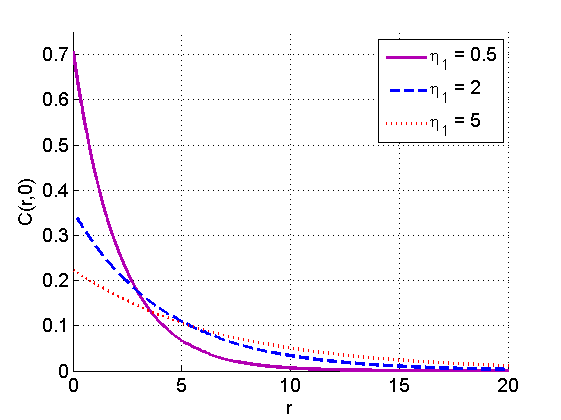}
\label{fig:ST cov d_1_t_0}
\end{subfigure}
\begin{subfigure}[]{}
\includegraphics[width=0.45\textwidth]{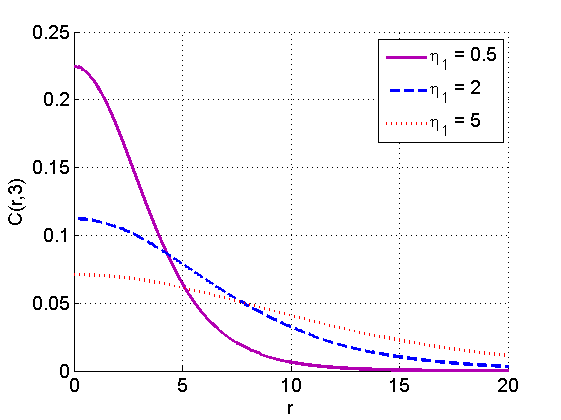}
\label{fig:ST cov d_1_t_3}
\end{subfigure}
\caption{Space-time covariances for $d=1, \eta_{0}=1, \xi=3, \tilde{D}=1 $. (a) Spatial dependence at $\tau=0$. 
(b) Spatial dependence at $\tau=3$.}\label{fig:1D_space_time_covariance_r}
\end{figure}

For $d=1,3$, $\mu=0$ it follows from \eqref{eq:fgc-spectral}
 that the corresponding covariances are given by the equations 

\begin{align}
\label{eq:fgc-spectral-d-1-m-0} 
C(\bfr,\tau)  = &
\frac{\E^{-\tilde{D}\,\vert \tau \vert\,}\eta_{0}}{\pi\,\e\,\xi} \, \int_{0}^{\infty} dk \,
\frac{\cos(kr)}{ \frac{1}{\e\,\xi^{2}} + k^2 }   \,\E^{-
\tilde{D}\,\vert \tau \vert\, \left(\eta_{1} k^{2} \xi^{2}
\right) }
\end{align}

and

\begin{align}
\label{eq:fgc-spectral-d-3-m-0} 
C(\bfr,\tau)  = &
\frac{\E^{-\tilde{D}\,\vert \tau \vert\,}\eta_0 \xi}{2\pi^{2}\e\,r} \, \int_{0}^{\infty} dk \,
\frac{k \,\sin(kr)}{\frac{1}{\e\,\xi^{2}} + k^2}  \,\E^{-
\tilde{D}\,\vert \tau \vert\, \left(\eta_{1} k^{2} \xi^{2}\right) }.
\end{align}

The above integrals can be evaluated analytically  for $\e>0$. 
Applying the identities [ Eq.(1.4.15), p.15]~\cite{magnus1954tables} 
and [ Eq.(2.4.26) p.74]~\cite{magnus1954tables}, the following explicit formulas 
 are derived for the one- and three-dimensional covariances 

%
%

\begin{align}
\label{eq:covariace-d-1-m-0} 
C(\bfr;\tau) & = \frac{\eta_{0}}{4\sqrt{\e}}\,\left[ \E^{-\frac{r}{\sqrt{\e}\xi}}
\erfc\left(\sqrt{\tilde{D}\,\vert \tau \vert}-\frac{r}{2\sqrt{\tilde{D}\e \,\vert \tau \vert\,}\xi}\right)\right.
\\\nonumber&
\left. \quad \quad \quad \quad + \, \E^{\frac{r}{\sqrt{\e}\xi}} 
\erfc\left(\sqrt{\tilde{D}\,\vert \tau \vert}+\frac{r}{2\sqrt{\tilde{D}\e \,\vert \tau \vert\,}\xi}\right)\right], 
\end{align}
 and 
 \begin{align}
 \label{eq:covariace-d-3-m-0} 
C(\bfr;\tau) & = \frac{\eta_0 \xi}{8\pi\e\,r}\,
\left[ \E^{-\frac{r}{\sqrt{\e}\xi}}  \erfc\left(\sqrt{\tilde{D}\,\vert \tau \vert}-
\frac{r}{2\sqrt{\tilde{D}\e \,\vert \tau \vert\,}\xi}\right) \right.
\\\nonumber&
\left. \quad \quad \quad \quad  -\E^{\frac{r}{\sqrt{\e}\xi}}\, \erfc\left(\sqrt{\tilde{D}\,\vert \tau \vert} + 
\frac{r}{2\sqrt{\tilde{D}\e \,\vert \tau \vert\,}\xi}\right)\right].
\end{align}

 The covariance function~\eqref{eq:covariace-d-1-m-0} is equivalent to a model  
 derived by Heine starting from the 
 general form of a second-order stochastic partial differential equation~\cite{heine1955models,jones1997models}.   
 Our result, however, is expressed in terms of  the parameters $\eta_0, \e, \xi, \tilde{D}$ that can be easily identified with properties of the 
 STRF and include the SSRF parameters of spatial random fields.   
 In particular, $\eta_0$ is an overall scale coefficient that determines the amplitude of the fluctuations and has units 
 $[X]^2$, $\e$ is a dimensionless 
 rigidity coefficient that suppresses large  gradients, $\xi$ is a characteristic length, 
 and $\tilde{D}$ is a characteristic inverse time. 
  We also derive the covariance function~\eqref{eq:covariace-d-3-m-0} 
 which is valid in three spatial dimensions.  Note that the presence of $\e$ in addition to $\xi$ implies that the 
 correlation length is a function of both parameters~\cite{dth11}. 
 In both $d=1$ and $d=3$ for  $\bfr\rightarrow 0$ and $\tau\rightarrow0$, the purely temporal and purely spatial 
 covariance models, respectively~\eqref{eq:cov-ssrf-zero-s2} 
 and~\eqref{eq:cov-ssrf-zero-t2}, are recovered from~\eqref{eq:covariace-d-1-m-0} and~\eqref{eq:covariace-d-3-m-0}. 
 In agreement with the zero-lag analysis, 
 the singularity at $r = 0$ and $\tau=0$ is present for $d=3$ in~\eqref{eq:covariace-d-3-m-0}. 
 This singularity, which denotes that the STRF has infinite variance, 
  is known in Gaussian field theory as well as in statistical modelling~\cite{jones1997models}. 
 We discuss a remedy  for this problem in the next section. 

In Figure~\ref{fig:ST cov} we plot the covariance functions~\eqref{eq:covariace-d-1-m-0} 
and~\eqref{eq:covariace-d-3-m-0} for $d=1$  and $d=3$, respectively.  For $d=3$ the minimum lags used 
are equal to $0.1$ to avoid the zero-lag singularity. 
Figures~\ref{fig:1D_space_time_covariance_t} and~\ref{fig:1D_space_time_covariance_r} 
demonstrate the dependence of the  covariance function~\eqref{eq:covariace-d-1-m-0}  in $d=1$
versus the spatial (time) lag under constant time (space) lag  
for different values of the rigidity coefficient $\e$. Note that at the larger lag distances, 
$r=3$ and $\tau=3$  the covariance function is smooth at the origin, i.e., for 
$\tau=0$ and $r=0$, respectively, in contrast with the non-differentiable peak at $(r=0, \tau=0)$.

\subsection{Small-$\mu$ approximation}

In the following we assume that $\mu \ll 1$, so that the curvature term in~\eqref{eq:ssrf-st} is present but 
multiplied with a small coefficient. This small perturbation, however, is sufficient to tame the 
singularity at the origin (zero lag) of the 
covariance function, effectively by reducing the impact of large-frequency fluctuations. 

%
%
%
%
We use the spectral representation~\eqref{eq:ssrf-st} for the covariance function. 
The exponential $\E^{-A(\ka, k)}$ in this case is expressed as follows

\[
\E^{-A(\ka, k)} = \E^{ - \ka \beta_{0} - u^{2}\, k^{2}   - \mu \, v\, k^4},
\]
where the coefficients $v$ and $u^2$ are defined by means of the following relations and 
the constant coefficients given by~\eqref{eq:cov-ssrf-st-2}
\beq
\label{eq:v-u2}
v = \xi^4 \left( \tilde{D}  \vert \tau \vert + \ka \beta_{0} \right),  \; 
u^{2}  = \eta_{1} \xi^{2}\,  \left(\tilde{D}\,\vert \tau \vert + \ka \beta_{0}\right).
\eeq 
The exponential component $\exp(-\mu \, v\, k^4)$ can be approximated using the Taylor expansion truncated 
at an even order $2M$ as follows
\[
\exp\left( - \mu \, v\, k^4 \right) \approx
\sum_{m=0}^{2M} \frac{(-1)^{m} \,\left(  \mu \, v\,  k^4\right)^m}{m!}.
\]
The even truncation order ensure that the approximation is  
non-negative. The divergence of the expansion for large $k$ is controlled by the 
exponential $\exp(-u\, k^2) $ for $\e >0$. Therefore, the truncated spectral density 
corresponds to a permissible covariance function.  The resulting expansion of the covariance 
function~\eqref{eq:ssrf-st} is given by the following series of integrals 
\begin{align}
\label{eq:cor-mod-C} 
C(\bfr,\tau)  \approx &
\frac{r^{1-d/2}\,\,\E^{-\tilde{D}\,\vert \tau \vert\,}}{(2\pi)^{d/2}} \, 
\sum_{m=0}^{2M } \frac{(-\mu )^m}{m!}\, \int_{0}^{\infty} d\ka \, v^{m}\,Q_{m} (\ka), 
\\
 Q_{m} (\ka) & = \E^{- \ka \beta_{0}}\int_{0}^{\infty} dk \, 
 k^{d/2+4m} \, J_{d/2-1}(kr)  \,\E^{- u^2\, k^{2}  }.
\end{align}


The integral over $k$ in $Q_{m} (\ka)$ can be evaluated using~\eqref{eq:spec-ssrf-int} which leads to 
\beq
\label{eq:Qmk}
Q_{m} (\ka) = \frac{\E^{- \ka \beta_{0}} }{u^{4m+d}}\, 
\left( \frac{r}{2} \right)^{d/2-1} \, \frac{\Gamma(d/2+2m)}{\Gamma(d/2)} \,
  {}_{1}F_{1}\left(2m+\frac{d}{2},\frac{d}{2};\frac{-r^2}{4\,u^2}\right).
\eeq

Based on the two equations above, the space-time covariance function becomes 
\begin{align}
\label{eq:cor-mod-D} 
C(\bfr,\tau)  \approx &
\frac{2\,\E^{-\tilde{D}\,\vert \tau \vert\,}}{(4\pi)^{d/2}} \, 
\sum_{m=0}^{2M } \frac{(-\mu )^m}{m!}\,\frac{\Gamma(d/2+2m)}{\Gamma(d/2)} \, R_{m}(r, \tau ),
\\
R_{m}(r,\tau) = & \int_{0}^{\infty} d\ka \, \frac{\E^{- \ka \beta_{0}} \, v^m}{u^{4m+d}}\, 
  {}_{1}F_{1}\left(2m+\frac{d}{2},\frac{d}{2};\frac{-r^2}{4\,u^2}\right).
\end{align}
In the above, the coefficients $v$ and $u^2$ are defined in~\eqref{eq:v-u2}, whereas the 
coefficients $\beta_{0}, \beta_{1}, \beta_{2}$ are defined in~\eqref{eq:cov-ssrf-st-2}.

\subsection{Numerical solutions for general values of $\mu$ and $\e$}

The presence of the parameter $\mu$  in the  general space-time model \eqref{eq:fgc-spectral} 
eliminates the observed singularities at the origin for $\mu=0$ . 
For $\mu>0$ and negative values of the rigidity coefficient $\e$, i.e. for $-2<\e<0$ 
the covariance function develops oscillations. 

Based on~\eqref{eq:fgc-spectral}, the 
space-time covariances for  $\mu \neq 0$ in $d=1,3$ are given by the following one dimensional integrals 
\begin{align}
\label{eq:fgc-spectral-d-1-all-m} 
C(\bfr,\tau)  = &
\frac{\E^{-\tilde{D}\,\vert\tau\vert}\eta_{0}\xi}{\pi} \, \int_{0}^{\infty} dk \,
\frac{\cos(kr)}{ 1 + \e\,k^2\xi^{2}+ \mu\,k^4\xi^{4} }   \,\E^{-
\tilde{D}\,\vert\tau\vert\, \left(\eta_{1} k^{2} \xi^{2}+ \mu\,k^4\xi^{4}
\right) },
\end{align}


\begin{align}
\label{eq:fgc-spectral-d-3-all-m-ns} 
C(\bfr,\tau)  = &
\frac{\E^{-\tilde{D}\,\vert\tau\vert}\eta_0 \xi^{3}}{2\pi^{2}\e} \, \int_{0}^{\infty} dk \,
\frac{k^2 \,\sinc(kr)}{1 + \e\,k^{2}\xi^{2}+ \mu\,k^{4}\xi^{4}}  \,\E^{-
\tilde{D}\,\vert\tau\vert\, \left(\eta_{1} k^{2} \xi^{2}+ \mu\,k^4\xi^{4}\right) }.
\end{align}

\begin{figure}
\centering
\begin{subfigure}[]{}
\includegraphics[width=0.45\textwidth]{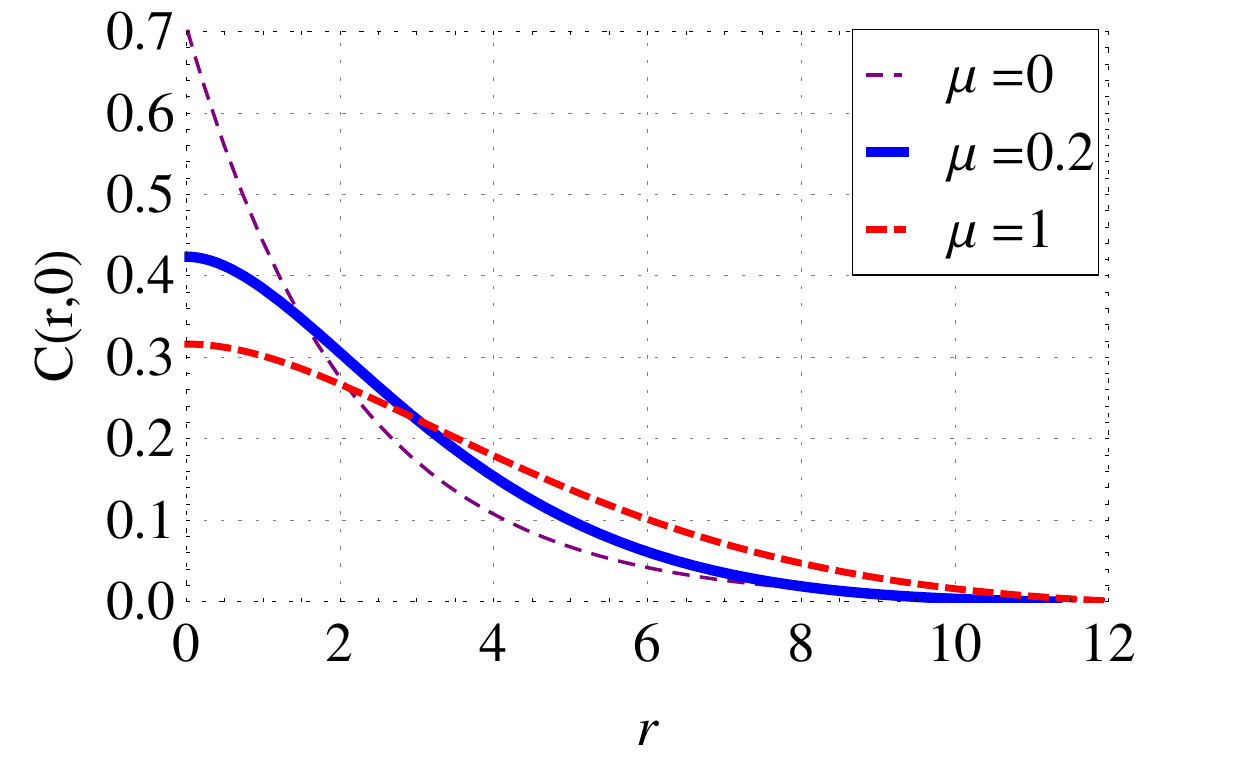}
\label{fig:ST cov d_1_t_0_m}
\end{subfigure}
\begin{subfigure}[]{}
\includegraphics[width=0.45\textwidth]{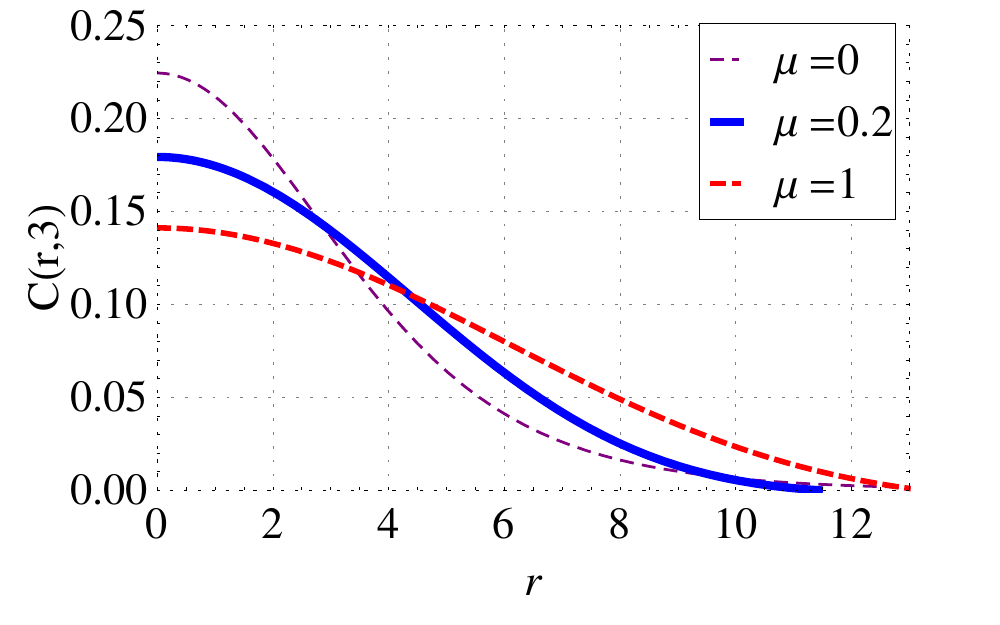}
\label{fig:ST cov d_1_t_3_m}
\end{subfigure}
\caption{Space-time covariances for $d=1, \eta_{0}=1,  \e=0.5, \xi=3, \tilde{D}=1 $. (a) Spatial dependence at $\tau=0$. 
(b) Spatial dependence at $\tau=3$.}\label{fig:1D_space_time_covariance_r_m}
\end{figure}

\begin{figure}
\centering
\begin{subfigure}[]{}
\includegraphics[width=0.45\textwidth]{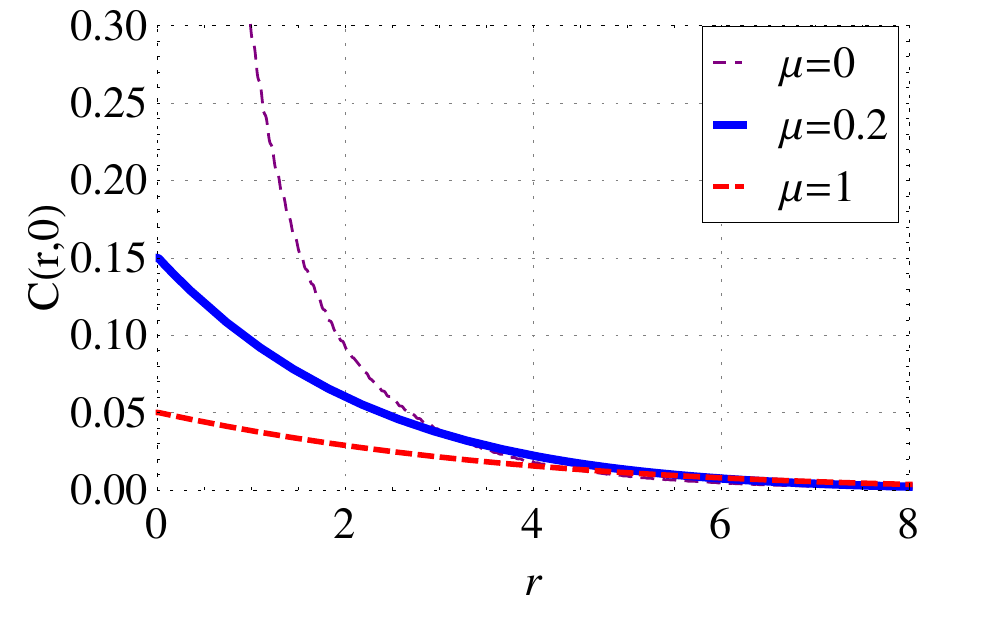}
\label{fig:ST cov d_3_t_0_m}
\end{subfigure}
\begin{subfigure}[]{}
\includegraphics[width=0.45\textwidth]{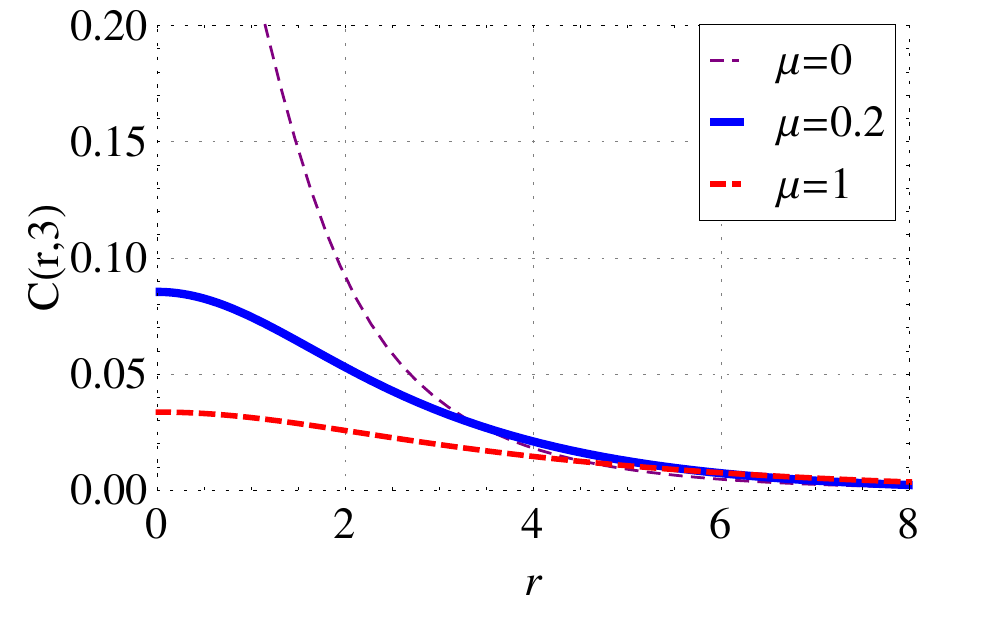}
\label{fig:ST cov d_3_t_3_m}
\end{subfigure}
\caption{Space-time covariances for $d=3,\e=0.5, \eta_{0}=1, \xi=3, \tilde{D}=1 $. (a) Spatial dependence at $\tau=0$. 
(b) Spatial dependence at $\tau=3$.}\label{fig:3D_space_time_covariance_r_m}
\end{figure}

The above integrals were evaluated in Mathematica using the  \verb+NIntegrate+ function with high working precision. 
The spectral integrals were truncated using a finite spectral cutoff, i.e. $\km=100$. 
In Figures~\ref{fig:1D_space_time_covariance_r_m} and~\ref{fig:3D_space_time_covariance_r_m} 
we plot the spatial dependence of the covariance functions~\eqref{eq:fgc-spectral-d-1-all-m} 
and~\eqref{eq:fgc-spectral-d-3-all-m-ns} for $d=1$  and $d=3$, respectively. 
Plots correspond to different values of  $\mu$ and  two different time lags. 
In all cases, the correlation length drops with increasing $\mu$. 
Moreover, it is evident in Fig~\ref{fig:3D_space_time_covariance_r_m} 
 that in the $d=3$ model has finite values at the origin for $\mu>0$.  

\begin{figure}
\centering
\begin{subfigure}[]{}
\includegraphics[width=0.45\textwidth]{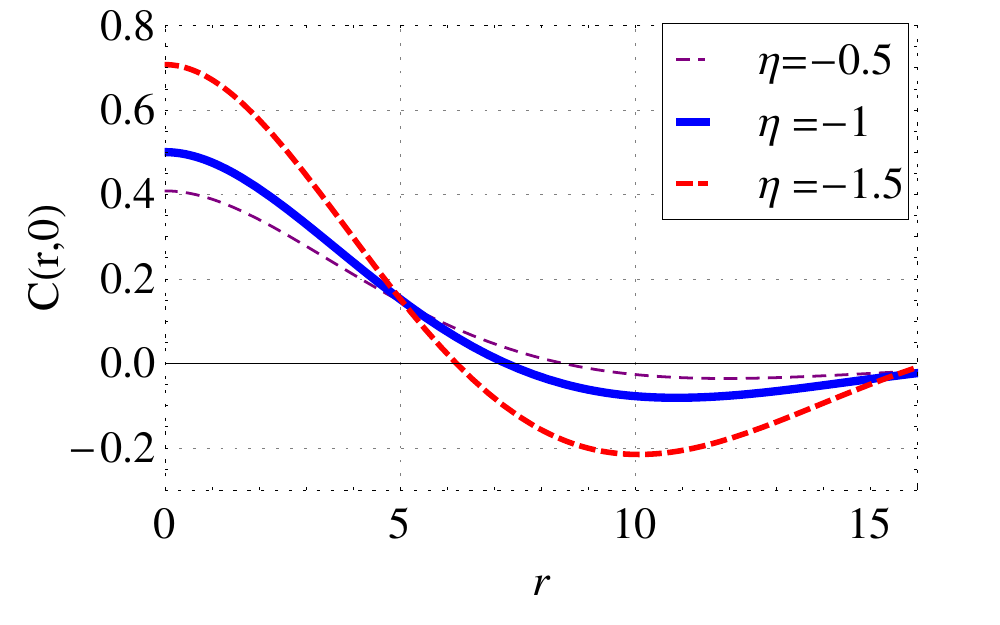}
\label{fig:ST cov d_1_t_0_nn}
\end{subfigure}
\begin{subfigure}[]{}
\includegraphics[width=0.45\textwidth]{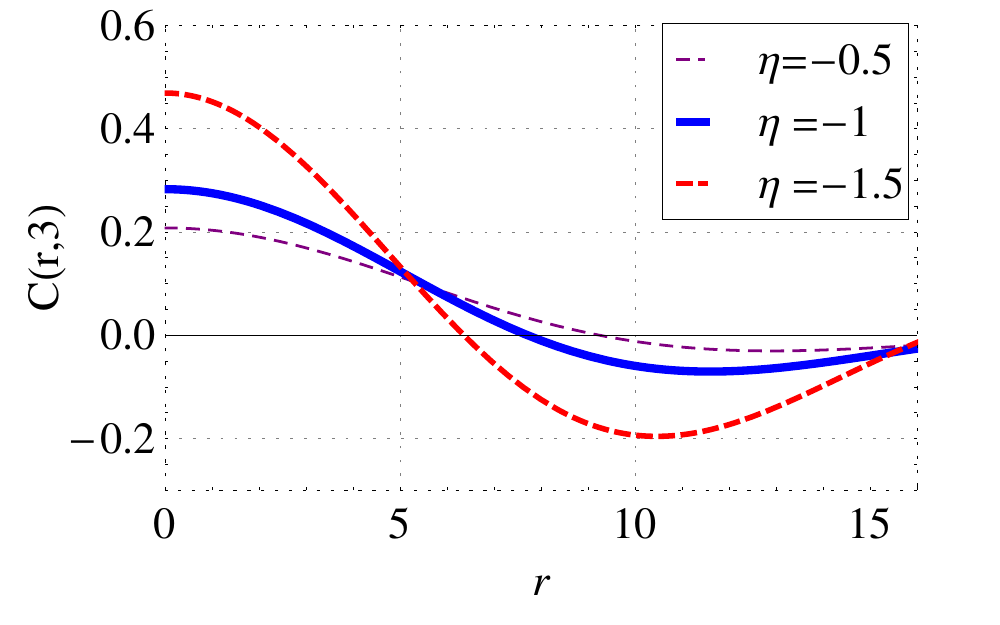}
\label{fig:ST cov d_1_t_3_nn}
\end{subfigure}
\caption{Oscillating space-time covariances for $d=1, \mu=1, \xi=3, \tilde{D}=1 $. (a) Spatial dependence at $\tau=0$. 
(b) Spatial dependence at $\tau=3$.}\label{fig:1D_space_time_covariance_r_nn}
\end{figure}

Figure~\ref{fig:1D_space_time_covariance_r_nn} 
displays how the covariance function~\eqref{eq:covariace-d-1-m-0}  
varies in $d=1$ with $r$ for  constant $\tau$ and  three  negative, 
values of $\e$. In all cases, the covariances obtained exhibit oscillations;  
such functions find applications in hydrology and in wave phenomena~\cite{dth14}.
In $d=3$ for negative values of $\e$  oscillating covariances are obtained by~\eqref{eq:fgc-spectral-d-3-all-m-ns};
 in this case,  however, the negative hole of the oscillations is reduced.

\section{Conclusions}
\label{sec:concl}
The analysis of spatiotemporal data and statistical mechanics have followed to a large extent 
separate paths,
in spite of the   conceptual overlaps between the two fields. 
The connection between statistical mechanics and geostatistics has 
been investigated in a series of papers~\cite{dth03,dth06,dthsel07,dthsel09,prem14,dth14}. 
Herein we demonstrate that space-time covariance functions for the analysis of spatiotemporal data sets  
can be obtained using ideas from statistical mechanics, such as field theory, maximum entropy, 
and linear response theory. 

We show that a Gaussian field model with an energy functional to which a curvature term is added 
can be derived from the principle of maximum entropy using suitable data-based constraints. 
The spatial covariance functions obtained from this energy functional incorporate 
``interactions'' between the sample locations, and they offer  increased flexibility due to
 a  richer parametric structure~\cite{dthsel09,dth11}.  
The ``equilibrium'' spatial  model is herein extended to the space-time domain by means 
of the covariance equation of motion or the associated Langevin 
equation for the realizations, both of which are obtained within the framework of the relaxation approximation 
in linear response theory. 
 The covariance equation of motion provides explicit formulas 
for the temporal evolution of the spectral density. It is also shown that
 the associated covariance functions in real space involve integrals 
 that can be analytically computed in different approximation regimes. 
 %
 The obtained solutions provide new types of non-separable, space-time correlation functions 
 without using simplifying but unrealistic 
 assumptions. In contrast with purely statistical models, the STRF model parameters 
 introduced in this paper correspond to  
 distinct properties of the random field. 
 
 For future investigations, flexible representations such as the Karhunen-Lo\`{e}ve  
 expansion can be used in combination with the Langevin equations 
 to obtain compact expressions for space-time covariance models. 
 The general formulation for  construction of space-time covariances based on concepts 
 of statistical mechanics 
 can be extended to incorporate more general models described by non-linear generalized Langevin equations 
 and colored or non-Gaussian driving noises \cite{athanassoulis2013two,Ivi2014two}.

\section*{Acknowledgment}
The research presented in this manuscript was funded by the project SPARTA 1591: ``Development of
Space-Time Random Fields based on Local Interaction Models and Applications in the Processing of
Spatiotemporal Datasets''. The project SPARTA is implemented under the ``ARISTEIA'' Action of the
 operational programme ``Education and Lifelong Learning'' and is co-funded by the European Social Fund
 and National Resources.

\vspace*{3pt}

\bibliographystyle{abbrv}
\bibliography{SpaceTime}

\begin{thebibliography}{10}

\bibitem{Porcu12}
In E.~Porcu, J.~Montero, and M.~Schlather, editors, {\textit {Advances and
  Challenges in Space-time Modelling of Natural Events}}, Lecture Notes in
  Statistics. Springer Berlin Heidelberg, 2012.

\bibitem{AbraSte}
M.~Abramowitz and I.~A. Stegun.
\newblock {\textit {Handbook of Mathematical Functions with Formulas, Graphs, and
  Mathematical Tables}}.
\newblock Dover, New York, 9th {D}over printing edition, 1964.

\bibitem{athanassoulis2013two}
G.~A. Athanassoulis, I.~C. Tsantili, and Z.~G. Kapelonis.
\newblock Two-time, response-excitation moment equations for a cubic
  half-oscillator under gaussian and cubic-gaussian colored excitation. Part 1:
  The monostable case.
\newblock {\textit{arXiv preprint arXiv:1304.2195}}, 2013.

\bibitem{Bochner59}
S.~Bochner.
\newblock {\textit {Lectures on Fourier Integrals}}.
\newblock Princeton University Press, Princeton, NJ, 1959.

\bibitem{Christakos92}
G.~Christakos.
\newblock {\textit {Random Field Models in Earth Sciences}}.
\newblock Academic Press, San Diego, 1992.

\bibitem{Cressie93}
N.~Cressie.
\newblock {\textit {Spatial Statistics}}.
\newblock John Wiley and Sons, New York, 1993.

\bibitem{Cressie11}
N.~Cressie and C.~L. Wikle.
\newblock {\textit {Statistics for Spatio-temporal Data}}.
\newblock John Wiley and Sons, New York, 2011.

\bibitem{deIaco02}
S.~De~Iaco, D.~Myers, and D.~Posa.
\newblock Nonseparable space-time covariance models: some parameteric families.
\newblock {\textit {Mathematical Geology}}, 34(1):23--42, 2002.

\bibitem{deIaco13}
S.~De~Iaco, D.~Posa, and D.~Myers.
\newblock Characteristics of some classes of space–time covariance functions.
\newblock {\textit {Journal of Statistical Planning and Inference}}, 143(11):2002 --
  2015, 2013.

\bibitem{dthetal08}
S.~N. Elogne, D.~Hristopulos, and E.~Varouchakis.
\newblock An application of {S}partan spatial random fields in environmental
  mapping: focus on automatic mapping capabilities.
\newblock {\textit {Serra}}, 22(5):633--646, 2008.

\bibitem{eldth08}
S.~N. Elogne and D.~T. Hristopulos.
\newblock Geostatistical applications of {S}partan spatial random fields.
\newblock In A.~Soares, M.~J. Pereira, and R.~Dimitrakopoulos, editors, {\textit
  geoENV VI � Geostatistics for Environmental Applications}, volume~15 of
  {\textit {Quantitative Geology and Geostatistics}}, pages 477--488. Springer,
  Berlin, Gemany, 2008.

\bibitem{Farmer07}
C.~L. Farmer.
\newblock Bayesian field theory applied to scattered data interpolation and
  inverse problems.
\newblock In A.~Iske and J.~Levesley, editors, {\textit {Algorithms for
  Approximation}}, pages 147--166. Springer-Verlag, Heidelberg, 2007.

\bibitem{Feynman65}
R.~P. Feynman and A.~R. Hibbs.
\newblock {\textit {Quantum Mechanics and Path Integrals}}.
\newblock McGraw-Hill, New York, 1965.

\bibitem{Forster90}
D.~Forster.
\newblock {\textit {Hydrodynamic Fluctuations, Broken Symmetry, and Correlation
  Functions}}.
\newblock Addison-Wesley, Redwood City, Calif., 1990.

\bibitem{Fuentes05}
M.~Fuentes, L.~Chen, J.~M. Davis, and G.~M. Lackmann.
\newblock Modeling and predicting complex space�time structures and patterns
  of coastal wind fields.
\newblock {\textit {Environmetrics}}, 16(5):449--464, 2005.

\bibitem{Furutsu63}
K.~Furutsu.
\newblock On the statistical theory of electromagnetic waves in a fluctuating
  medium.
\newblock {\textit {Journal of Research of the National Institute of Standards and
  Technology}}, 67D(3):303--323, 1963.

\bibitem{Gelfand48}
I.~M. Gel'fand and A.~M. Yaglom.
\newblock Integration in functional spaces and its applications in quantum
  physics.
\newblock {\textit {Journal of Mathematical Physics}}, 1(1):48--69, 1960.

\bibitem{Goldenfeld92}
N.~Goldenfeld.
\newblock {\textit {Lectures on Phase Transitions and the Renormalization Group}}.
\newblock Frontiers in Physics, 85. Addison-Wesley, 1992.

\bibitem{Gradshteyn94}
I.~S. Gradshteyn and I.~M. Ryzhik.
\newblock {\textit {Tables of Integrals, Series, and Products}}.
\newblock Academic Press, New York, 5th edition, 1994.

\bibitem{Guttorp06}
P.~Guttorp and T.~Gneiting.
\newblock Studies in the history of probability and statistics xlix on the
  matern correlation family.
\newblock {\textit {Biometrika}}, 93(4):989--995, 2006.

\bibitem{heine1955models}
V.~Heine.
\newblock Models for two-dimensional stationary stochastic processes.
\newblock {\textit {Biometrika}}, 42(1-2):170--178, 1955.

\bibitem{Hoffman13}
M.~D. Hoffman, D.~M. Blei, C.~Wang, and J.~Paisley.
\newblock Stochastic variational inference.
\newblock {\textit {The Journal of Machine Learning Research}}, 14(1):1303--1347,
  2013.

\bibitem{Hohenberg77}
P.~C. Hohenberg and B.~I. Halperin.
\newblock Theory of dynamic critical phenomena.
\newblock {\textit {Review of Modern Physics}}, 49:435--479, 1977.

\bibitem{dthsel07}
D.~Hristopulos and S.~Elogne.
\newblock Analytic properties and covariance functions of a new class of
  generalized {G}ibbs random fields.
\newblock {\textit {IEEE Transactions on Information Theory}}, 53(12):4667--4679,
  2007.

\bibitem{dth03}
D.~T. Hristopulos.
\newblock {S}partan {G}ibbs random field models for geostatistical
  applications.
\newblock {\textit {SIAM Journal of Scientific Computing}}, 24(6):2125--2162, 2003.

\bibitem{dth06}
D.~T. Hristopulos.
\newblock Spatial random field models inspired from statistical physics with
  applications in the geosciences.
\newblock {\textit {Physica A: Statistical Mechanics and its Applications}},
  365(1):211--216, 2006.

\bibitem{dth14}
D.~T. Hristopulos.
\newblock Covariance functions motivated by spatial random field models with
  local interactions.
\newblock {\textit {Stochastic Environmental Research and Risk Assessment}},
  29:739--–754, 2015.

\bibitem{dthsel09}
D.~T. Hristopulos and S.~N. Elogne.
\newblock Computationally efficient spatial interpolators based on {S}partan
  spatial random fields.
\newblock {\textit {{IEEE} Transactions on Signal Processing}}, 57(9):3475--3487,
  2009.

\bibitem{prem14}
D.~T. Hristopulos and E.~Porcu.
\newblock Multivariate spartan spatial random field models.
\newblock {\textit {Probabilistic Engineering Mechanics}}, 37:84--92, 2014.

\bibitem{dth11}
D.~T. Hristopulos and M.~\v{Z}ukovi\v{c}.
\newblock Relationships between correlation lengths and integral scales for
  covariance models with more than two parameters.
\newblock {\textit {Stochastic Environmental Research and Risk Assessment}},
  25(1):11--19, 2011.

\bibitem{Itzykson89}
C.~Itzykson and J.-M. Drouffe.
\newblock {\textit {Statistical field theory}}, volume~1.
\newblock Cambridge University press, New York, 1989.

\bibitem{Jaynes57}
E.~T. Jaynes.
\newblock Information theory and statistical mechanics.
\newblock {\textit {Physical review}}, 106(4):620--630, 1957.

\bibitem{jones1997models}
R.~H. Jones and Y.~Zhang.
\newblock Models for continuous stationary space-time processes.
\newblock In {\textit {Modelling longitudinal and spatially correlated data}}, pages
  289--298. Springer, 1997.

\bibitem{Kalnay03}
E.~Kalnay.
\newblock {\textit {Atmospheric modeling, data assimilation, and predictability}}.
\newblock Cambridge, Cambridge, 2003.

\bibitem{Kardar07}
M.~Kardar.
\newblock {\textit {Statistical Physics of Fields}}.
\newblock Cambridge University Press, 2007.

\bibitem{Kolovos04}
A.~Kolovos, G.~Christakos, D.~Hristopulos, and M.~L. Serre.
\newblock Methods for generating non-separable spatiotemporal covariance models
  with potential environmental applications.
\newblock {\textit {Advances in Water Resources}}, 27(8):815--830, 2004.

\bibitem{Lemm05}
J.~C. Lemm.
\newblock {\textit {Bayesian Field Theory}}.
\newblock Johns Hopkins University Press, Baltimore, 2005.

\bibitem{Lim09}
S.~C. Lim and L.~P. Teo.
\newblock Generalized {W}hittle {M}at\'{e}rn random field as a model of
  correlated fluctuations.
\newblock {\textit {Journal of Physics A: Mathematical and Theoretical}},
  42:{105202}, 2009.

\bibitem{Ma05}
C.~Ma.
\newblock Linear combinations of space-time covariance functions and
  variograms.
\newblock {\textit {{IEEE} Transactions on Signal Processing}}, 53(3):857--864, 2005.

\bibitem{Ma08}
C.~Ma.
\newblock Recent developments on the construction of spatio-temporal covariance
  models.
\newblock {\textit {Stochastic Environmental Research and Risk Assessment}},
  22(1):S39--S47, 2008.

\bibitem{magnus1954tables}
W.~Magnus, H.~Bateman, A.~Erd{\'e}lyi, F.~Oberhettinger, and F.~Tricomi.
\newblock {\textit {Tables of integral transforms}}.
\newblock McGraw-Hill, 1954.

\bibitem{Marconi08}
U.~M.~B. Marconi, A.~Puglisi, L.~Rondonic, and A.~Vulpiani.
\newblock Fluctuation dissipation: Response theory in statistical physics.
\newblock {\textit {Physics Reports}}, 461(4):111--195, 2008.

\bibitem{Mussardo10}
G.~Mussardo.
\newblock {\textit {Statistical field theory}}.
\newblock Oxford Univ. Press, 2010.

\bibitem{Novikov65}
E.~A. Novikov.
\newblock Functionals and the random-force method in turbulence theory.
\newblock {\textit {Sov. Phys. JETP}}, 20(5):1290--�1294, 1965.

\bibitem{Rasmussen06}
C.~E. Rasmussen and C.~K.~I. Williams.
\newblock {\textit {Gaussian Processes for Machine Learning}}.
\newblock MIT Press, Massachusetts Institute of Technology, 2006.

\bibitem{Stein05}
M.~L. Stein.
\newblock Space–time covariance functions.
\newblock {\textit {Journal of the American Statistical Association}},
  100(469):310--321, 2005.

\bibitem{Swift77}
J.~Swift and P.~C. Hohenberg.
\newblock Hydrodynamic fluctuations at the convective instability.
\newblock {\textit {Physical Review A}}, 15(1):319, 1977.

\bibitem{Tarantola05}
A.~Tarantola.
\newblock {\textit {Inverse problem theory and methods for model parameter
  estimation}}.
\newblock SIAM, Philadelphia, 2005.

\bibitem{Ivi2014two}
I.~S.~C. Tsantili.
\newblock {\textit {Two-time response excitation theory for non-linear stochastic
  dynamical systems}}.
\newblock PhD thesis, National Technical University of Athens, 2013.

\bibitem{zuk08b}
M.~\v{Z}ukovi\v{c} and D.~T. Hristopulos.
\newblock Environmental time series interpolation based on {S}partan random
  processes.
\newblock {\textit {Atmospheric Environment}}, 42(33):7669--7678, 2008.

\bibitem{zuk09c}
M.~\v{Z}ukovi\v{c} and D.~T. Hristopulos.
\newblock The method of normalized correlations: A fast parameter estimation
  method for random processes and isotropic random fields that focuses on
  short-range dependence.
\newblock {\textit {Technometrics}}, 51(2):173--185, 2009.

\bibitem{Wackernagel03}
H.~Wackernagel.
\newblock {\textit {Multivariate Geostatistics}}.
\newblock Springer Verlag, Berlin, 3rd edition, 2003.

\bibitem{Whittle54}
P.~Whittle.
\newblock On stationary processes in the plane.
\newblock {\textit {Biometrika}}, 41(3/4):434--�449, 1954.

\bibitem{Xu05}
Q.~Xu.
\newblock Representations of inverse covariances by differential operators.
\newblock {\textit {Advances in Atmospheric Sciences}}, 22(2):181--198, 2005.

\bibitem{Yaglom87}
A.~M. Yaglom.
\newblock {\textit {Correlation Theory of Stationary and Related Random Functions
  I}}.
\newblock Springer Verlag, New York, 1987.

\bibitem{Yaremchuk12}
M.~Yaremchuk and A.~Sentchev.
\newblock Multi-scale correlation functions associated with polynomials of the
  diffusion operator.
\newblock {\textit {Quarterly Journal of the Royal Meteorological Society}},
  138(668):1948--1953, 2012.

\bibitem{yaremchuk2012multi}
M.~Yaremchuk and A.~Sentchev.
\newblock Multi-scale correlation functions associated with polynomials of the
  diffusion operator.
\newblock {\textit {Quarterly Journal of the Royal Meteorological Society}},
  138(668):1948--1953, 2012.

\bibitem{Yaremchuk11}
M.~Yaremchuk and S.~Smith.
\newblock On the correlation functions associated with polynomials of the
  diffusion operator.
\newblock {\textit {Quarterly Journal of the Royal Meteorological Society}},
  137(660):1927--1932, 2011.

\bibitem{Zinn04}
J.~Zinn-Justin.
\newblock {\textit {Quantum Field Theory and Critical Phenomena}}.
\newblock Oxford Univ. Press, Oxford, 4th edition, 2004.

\end{thebibliography}

\end{document}